\documentclass[aps,preprint,showpacs,preprintnumbers,superscriptaddress]{revtex4-1}
\usepackage{multirow}
\usepackage{diagbox}
\usepackage{booktabs}
\usepackage{array}
\usepackage{natbib}
\usepackage{amsmath,amssymb,graphicx,bm,subfigure,float,siunitx,color}
\usepackage{bm,subfigure,float,siunitx,graphicx}
\usepackage[figuresright]{rotating}
\usepackage{color}
\usepackage{epstopdf}
\usepackage{mathrsfs}

\begin{document}
\title{\boldmath 
Probing Einstein-Maxwell-Scalar Black hole via Thin Accretion Disks and Shadows with EHT Observations of M87* and Sgr A*}
\author{Yingdong Wu }
\email{Email address: yingdongwu7@gmail.com}
\affiliation{Department of Physics, State key laboratory of quantum functional materials, and Guangdong Basic Research Center of Excellence for Quantum Science, Southern University of Science and Technology, Shenzhen 518055, China}

\author{Ziqiang Cai}
\email{Email address:gs.zqcai24@gzu.edu.cn}
\affiliation{School of Physics, Guizhou University, Guiyang 550025, China}

\author{Zhenglong Ban }
\email{Email address:gs.zlban22@gzu.edu.cn}
\affiliation{School of Physics, Guizhou University, Guiyang 550025, China}

\author{Haiyuan Feng \footnote{Corresponding author}}
\email{Email address:  fenghaiyuanphysics@gmail.com}
\affiliation{Department of Physics, State key laboratory of quantum functional materials, and Guangdong Basic Research Center of Excellence for Quantum Science, Southern University of Science and Technology, Shenzhen 518055, China}

\affiliation{School of Physics and Electronic Engineering, Shanxi Normal University, Taiyuan 030031, China}

\author{Wei-Qiang Chen  }
\email{Email address: chenwq@sustech.edu.cn}
\affiliation{Department of Physics, State key laboratory of quantum functional materials, and Guangdong Basic Research Center of Excellence for Quantum Science, Southern University of Science and Technology, Shenzhen 518055, China}

\affiliation{Quantum Science Center of Guangdong-Hong Kong-Macao Greater Bay Area, Shenzhen 518045, China}

\begin{abstract}
We investigated the shadows and thin accretion disks of Einstein-Maxwell-Scalar (EMS) black hole. Firstly, we investigated the influence of EMS parameters on the black hole shadow using the null geodesic method and constrained these parameters based on EHT observations of M87* and Sgr A*. Furthermore, we analyzed the direct emission, lensing ring, and photon ring structures in EMS black hole. Comparing our results with the Schwarzschild and Reissner-Nordstr$\ddot{\mathrm{o}}$m (RN) black holes, we found that the Schwarzschild black hole exhibits the largest shadow radius and the highest observed intensity.

\end{abstract}

\maketitle
\section{Introduction}
Black holes (BHs) represent one of the fundamental solutions to Einstein’s field equations and are regarded as robust predictions of General Relativity (GR). While historically debated, numerous observational efforts have been made to confirm their existence. A significant breakthrough occurred in 2015 when the Laser Interferometer Gravitational-Wave Observatory (LIGO) detected gravitational waves (GWs) originating from a binary BH merger \cite{124}. Subsequently, in 2019, the Event Horizon Telescope (EHT) provided the first direct observational evidence of a BH shadow at the center of the galaxy M87  \cite{125,126,127,128,129,130}.

Further advancements in EHT observations led to the polarization imaging of M87*, revealing the presence of a magnetic field, which provides insights into jet formation mechanisms \cite{131,132,133}. Additionally, EHT reported the detection of the shadow of SgrA*, the supermassive BH at the center of the Milky Way \cite{134}. These observations offer valuable constraints on the geometry of spacetime in the strong-field regime, particularly near the event horizon, allowing precise measurements of BH mass and spin \cite{135,136,145,146,147,148,149,150,151,152,153,154,155,185}. Collectively, these findings provide a crucial platform for testing GR and exploring potential modifications to gravitational theories in extreme astrophysical environments \cite{137,Liu:2024brf,Yin:2025coq,He:2022lrc,Wu:2024sng,Feng:2024iqj,Feng:2023iha,Feng:2022bst}.  

It is well known that GR has demonstrated remarkable accuracy in various tests \cite{138}. However, it faces certain limitations within the cosmological framework. These include challenges in explaining the accelerated expansion of the universe, the presence of initial singularities, the missing mass problem, and the Coincidence Problem. Furthermore, testing GR in the strong-field regime, particularly in the vicinity of BH horizons, remains a significant challenge \cite{139}. In addition, modifications to GR have been explored to extend our understanding beyond its conventional framework, aiming to probe near-horizon physics and the large-scale structure of the universe \cite{140,141,142,180}.

Among various modified gravity models, the Einstein-Maxwell-Scalar (EMS) model serves as a promising candidate. In fact, various specific realizations of EMS models naturally arise as consistent truncations or low-energy limits of string theory.
For instance, in heterotic string theory, compactification from ten to four dimensions in the Einstein frame yields an effective action containing the metric, a dilaton-like scalar field, and Abelian gauge fields with an exponential coupling of the form $f(\phi)F_{\mu\nu}F^{\mu\nu}$ — as in the well-known Gibbons–Maeda–Garfinkle–Horowitz–Strominger (GMGHS) solution \cite{Gibbons:1988,Garfinkle:1990qj}. Similar structures occur in Type II string compactifications, where moduli fields (Kähler, complex structure, axion–dilaton) couple to gauge fields descending from the NS–NS and R–R sectors \cite{Polchinski:1998,Becker:2006}.
In extended supergravity, particularly $\mathcal{N}=4$ and $\mathcal{N}=8$ theories, the gauge kinetic matrix $f_{IJ}(\phi)$ depends explicitly on scalar fields \cite{Cremmer:1979,deWit:1982}. Truncation to a single $U(1)$ vector and one scalar degree of freedom yields precisely the EMS form. The $\mathcal{N}=2$ STU model \cite{Duff:1996} provides another example, in which selecting a single scalar–vector sector produces an EMS-type action.

The EMS action serves as a powerful unifying framework, encompassing two distinct and mutually exclusive branches of black hole solutions: the Einstein--Maxwell--Dilaton (EMD) branch and the spontaneously scalarized Reissner--Nordström (RN) branch. The profound differences between these two scenarios are encoded solely in the local behavior of the coupling function $K(\phi)$ at $\phi=0$. The EMD limit, realized by choosing an exponential coupling $K(\phi)=\mathrm{e}^{2\alpha\phi}$, is characterized by $K(0)=1$ and $K'(0)\neq0$. This condition renders the electro-vacuum RN solution, where $\phi\equiv0$, an invalid background. Consequently, EMD black holes are born with an intrinsic, non-trivial scalar field, often referred to as ``dilaton hair'' and they never bifurcate from a hairless RN seed. Their asymptotic behavior is distinct, approaching a dilaton asymptotic at infinity, and they can violate the RN extremal bound on the charge-to-mass ratio \cite{Garfinkle:1990qj,belkhadria_2025}.

In contrast, the spontaneously scalarized RN branch emerges from a different class of coupling functions satisfying $K(0)=1$, $K'(0)=0$, and $K''(0)<0$. This specific set of conditions permits the RN metric with $\phi\equiv0$ as a legitimate and stable background in the absence of a strong electric field. However, in the presence of a sufficiently strong electric charge, a tachyonic instability can arise. The effective mass squared of the scalar field, $\mu_{\text{eff}}^{2}(r)=K''(0)\,Q^{2}/r^{4}$, becomes negative within a certain radial window. This instability triggers a second-order phase transition, where the hairless RN black hole spontaneously acquires scalar hair, bifurcating from the original RN branch at a critical charge \cite{Qiu_2021}. Asymptotically, these scalarized black holes return to the RN solution, as the scalar hair vanishes at infinity.

%\textbf{The divergent nature of these two branches extends to their observable signatures, providing distinct avenues for observational tests. The EMD model predicts a gyromagnetic ratio for black holes that is consistently less than two ($g<2$). For scalarized RN black holes, the gyromagnetic ratio interpolates between $g=2$ for the hairless solution and lower values as the scalar charge increases \cite{belkhadria_2025}. }

Since the EMS theory allows for hairy BH solutions \cite{143,144,145,189}, where BHs can possess scalar hair, investigating their properties in the strong-field regime is of significant interest. Previous studies have shown that dilaton BH spacetimes are influenced by the dilaton charge, which not only serves as a new form of hair but also provides a powerful tool for analyzing optical phenomena in such backgrounds. In this work, we consider a charged BH solution within the EMS theory as an extension of the RN solution. Unlike dilaton BHs, where the scalar field typically contributes through an effective charge, the BH solution in EMS theory is directly governed by its model parameters, which serve as the primary distinguishing feature.
A comprehensive exploration of the physical and optical properties of this solution is therefore essential. Investigating the effects of its parameters on spacetime geometry and optical phenomena will contribute to a deeper understanding of its astrophysical implications and provide a means to differentiate it from other BH solutions.

In this article, we analyzed the optical properties of this BH solution through its shadow and the observable appearance of accretion disk emission in the strong-field regime. We further investigated the impact of EMS BH parameters on optical phenomena and constrained the viable ranges of parameters \( \alpha \) and \( \beta \) using EHT observational data. Our paper is organized as follows. In Sec.II, we will give the BH solution in EMS theory. In Sec.III, we will discuss the influence of EMS parameters on the optical properties of BH and plot the photon orbits of BH. Furthermore, we investigated the observable effects of radiation emitted by the accretion disk and found that the EMS BH exhibits a significantly reduced radiation intensity compared to the Schwarzschild and RN BHs. And in Sec III.E, we also explore the deflection angle in the weak  gravitational lensing. Finally, we present our conclusions and provide an outlook in Sec.IV. The paper has been written with the unit system $G=M=c=1$, and metric signature $(-,+,+,+)$.

\section{The black hole solution in Einstein-Maxwell-scalar theory}	
The systematic investigation of scalar-gravity coupling was introduced by Fisher, who obtained static spherically symmetric solutions to Einstein's equations with a massless scalar field \cite{102}. This foundational work 
established the theoretical basis for subsequent developments, particularly in 
Einstein-Maxwell-scalar (EMS) theory which has attracted significant attention 
due to its implications and natural emergence in: (i) Kaluza-Klein 
dimensional reduction \cite{103}, (ii) dilaton couplings in string theory 
\cite{104}, and (iii) cosmological inflation and late-time acceleration 
\cite{105}.

% The scalar-electromagnetic coupling, arising naturally from dimensional 
% reduction, is typically described by the interaction term:
% \begin{equation}
% \mathcal{L}{\text{int}} = f(\phi) F{\mu\nu} F^{\mu\nu}
% \end{equation}
% where $f(\phi)$ governs the coupling form (constant/exponential/other 
% functional forms). Such coupling modifies electromagnetic propagation in 
% strong-field regimes (e.g., near black holes) and cosmological scales, while 
% introducing non-trivial corrections to Einstein's field equations. Notably, EMS 
% theory predicts novel black hole solutions with scalar-electromagnetic 
% coupling, potentially yielding detectable signatures in astrophysical 
% observations and laboratory experiments \cite{169,170,171,172,173,174,175,186}.
%The EMS theory—a low-energy effective theory in string theory—generalizes Einstein-Maxwell theory by incorporating a dynamical scalar field coupled to the Maxwell term \cite{187,188}. 
The EMS action takes the form
\begin{equation}
\label{1}
S = \frac{1}{16\pi}\int d^4x\sqrt{-g}\left[R - 2(\nabla\phi)^2 - K(\phi)F^2 - V(\phi)\right],
\end{equation}
where $R$ is the Ricci scalar, $\phi$ is the scalar field. $F_{\mu\nu}=\partial_{\mu}A_{\nu}-\partial_{\nu}A_{\mu}$ represents the electro-magnetic field tensor. The field equations derived from this action yield
\begin{equation}
\begin{cases}
\nabla_\mu[K(\phi)F^{\mu\nu}] = 0, \\
\Box\phi = \frac{1}{4}\left[\partial_\phi K(\phi)F^2+\partial_\phi V(\phi) \right], \\
R_{\mu\nu} = 2\partial_\mu\phi\partial_\nu\phi + \frac{1}{2}g_{\mu\nu}V(\phi) +2K(\phi)(F_{\mu\sigma}F^\sigma_{~\nu} - \frac{1}{4}g_{\mu\nu}F^2).
\end{cases}
\end{equation}

The coupling function $K(\phi)$ and potential $V(\phi)$ determine the theory's physical content. Some important solutions with different $K(\phi)$ and $V(\phi)$ are given by \cite{3,4,5,6,7,8}. Notable special solutions include
\begin{itemize}
\item RN-de Sitter BH :$K(\phi)=1$, $V=2\Lambda$, $\phi=0$,
\item Dilaton BH :$K(\phi)=e^{2\phi}$, $V=0$ \cite{1,2}.
\end{itemize}

Different choices of the coupling function $K(\phi)$ lead to qualitatively distinct physics. Polynomial forms (such as $K(\phi)=1+\alpha\phi^2$) preserve the RN branch but allow for spontaneous scalarization above a critical threshold, where the scalar field effectively screens the electromagnetic charge and produces multiple scalarized branches. Exponential couplings (such as $K(\phi)=e^{2\alpha\phi}$) correspond to dilatonic models in which RN solutions no longer exist and charged black holes necessarily carry scalar hair, with modified extremality and thermodynamics. Other bounded or interpolating forms regulate the growth of scalar hair or bridge between EMS-type and dilatonic behavior.

Our investigation focuses on the parameterized coupling function \cite{Yu_2021}
\begin{equation}
\label{5}
K(\phi) = \frac{(\alpha^2+1)e^{-2\phi/\alpha}}{(\alpha^2+1+\beta)e^{-2\phi(\alpha^2+1)/\alpha} + \beta\alpha^2},
\end{equation}
which exhibits two physically significant limits. (i) $\beta\to\infty$: Decoupled electromagnetic sector. (ii) $\beta\to 0$: Einstein-Maxwell-Dilaton (EMD) theory with $K(\phi)=e^{2\alpha\phi}$. (iii) $\beta\to 0$: Einstein-Maxwell-Dilaton (EMD) theory with $K(\phi)=e^{2\alpha\phi}$.

The physical meaning of $\alpha$ is that it measures the strength of the coupling between the dilaton scalar field and the Maxwell field. It first emerges in the EMD limit ($\beta=0$) and is inherited from low-energy string theory, where the dilaton couples universally to gauge fields. Parameter $\beta$ controls the relative contribution of the RN and dilaton branches in the metric. Regarding its origin, it is introduced to combine the RN and dilaton line elements into a single interpolating family, thereby extending the EMD class.

As shown in Ref.\cite{9}, these admit BH solution with metric components
\begin{equation}
\begin{cases}
ds^2 = -f(r)dt^2 + f^{-1}(r)dr^2 + C(r)d\Omega^2, \\
f(r) = \left(1-\frac{b_1}{r}\right)\left(1-\frac{b_2}{r}\right)^{\frac{1-\alpha^2}{1+\alpha^2}} + \frac{\beta Q^2}{C(r)}, \\
C(r) = r^2\left(1-\frac{b_2}{r}\right)^{\frac{2\alpha^2}{1+\alpha^2}},
\end{cases}
\end{equation}
where the parameters $b_1$ and $b_2$ are determined by
\begin{equation}
\begin{cases}
b_1 = \left(1 + \sqrt{1 - q^2(1-\alpha^2)}\right)M ,\\
b_2 = \frac{1+\alpha^2}{1-\alpha^2}\left[1 - \sqrt{1 - q^2(1-\alpha^2)}\right]M,
\end{cases}
\end{equation}
with $q\equiv Q/M$ representing the dimensionless charge-to-mass ratio. Moreover, since obtaining analytic expressions for the inner and outer horizons of the EMS black hole is extremely difficult, we consider two special cases, while the general EMS case is determined numerically. (I) $\alpha=1$: With this definition established, the following relations can be derived \cite{Yu_2021}
\begin{equation}
\begin{cases}
r_{+}=M\sqrt{2-q^2-\beta q^2+\sqrt{\left(2-q^2\right)^2-4q^2\beta}},\\
r_{-}=M\sqrt{2-q^2-\beta q^2-\sqrt{\left(2-q^2\right)^2-4 q^2\beta}}.
\end{cases}
\end{equation}
For analytical convenience, we take  $Q > 0$  without imposing any loss of generality. The condition ensuring the existence of both the inner and outer horizons can be demonstrated to take the form
\begin{eqnarray}
0<q<\sqrt{2},\quad \quad \beta<\frac{\left(2-q^2\right)^2}{4q^2}.
\end{eqnarray}
Moreover, the condition for the existence of a single horizon can be written as
\begin{eqnarray}
0<q<\sqrt{2}, \quad  \quad \beta=\frac{(2-q^2)^2}{4q^2},
\end{eqnarray}
under which the two horizons coincide, giving rise to an extremal black hole. This demonstrates that an extremal configuration can occur once the coupling constant $\beta$ meets the above criterion, irrespective of how small the charge is. (2) $\beta=0$ (the EMD solution): In this case, the inner and outer horizons are determined by $b_{1}$ and $b_{2}$.
\begin{equation}
\begin{cases}
r_{+}=b_1 \text { (the outer event horizon) }, \\
r_{-}=b_2 \text { (the inner Cauchy horizon) }.
\end{cases}
\end{equation}
The horizons are determined by the black hole's mass $M$, charge $Q$, and the dilaton coupling constant $\alpha$. For two distinct real horizons to exist, the condition $q^2 \leq 1+\alpha^2$ must hold. The equality $q^2 = 1+\alpha^2$ corresponds to the extremal case, where the two horizons merge into a single degenerate horizon. In contrast, the condition $q^2 > 1+\alpha^2$ eliminates the horizons and exposes a naked singularity. Therefore, when $\alpha \neq 1$ with arbitrary $\beta$, or when $\beta = 0$ with arbitrary $\alpha$, a case-by-case analysis is necessary to ensure that naked singularities do not emerge in the black hole solutions. Notably, both situations correspond to special subclasses of the EMS framework. In the following analysis, we confine the parameters to these well-defined ranges to guarantee the existence of black hole solution.

Besides these, the black-hole horizons for all other parameter choices are already plotted in Figure 1. In the figure below, we plot the variation of the metric components of the EMS solution with respect to the coordinate radius $r$.
\begin{figure}[H]
	\begin{minipage}{0.5\textwidth}
  		\includegraphics[scale=0.9,angle=0]{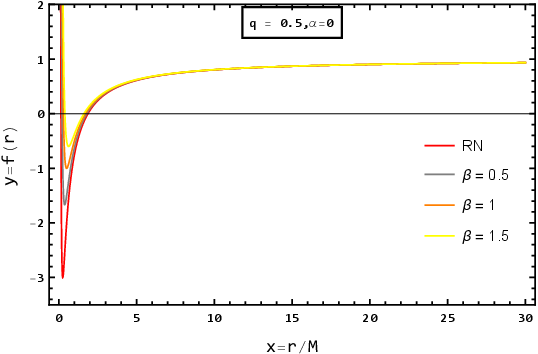}
	\end{minipage}
    \begin{minipage}{0.5\textwidth}
		\includegraphics[scale=0.9,angle=0]{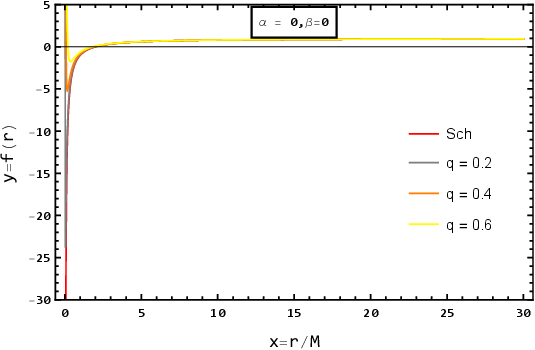}
	\end{minipage}	    
    \begin{minipage}{0.5\textwidth}
		\includegraphics[scale=0.9,angle=0]{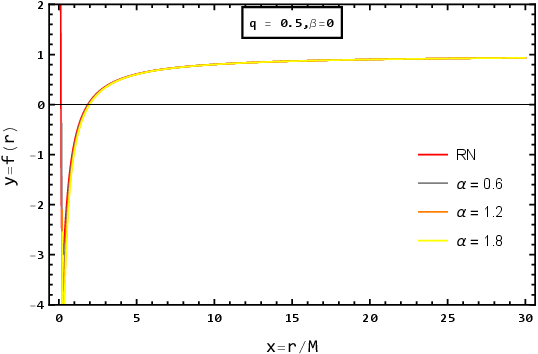}
	\end{minipage}
\caption{\label{Fig.1}The above three pictures plot the function $f(r)$  for different values of $\beta$,$\alpha$ and $q$. } 
\end{figure}
From the above two pictures, we can see there are two intersection points with $y=0$ which means the existence of the inner and outer horizons of BH. However, in the third case (EMD), we observe that both a single intersection and two intersections can occur. All such parameter choices lie within the admissible ranges discussed above, thereby ensuring the absence of naked singularities. Furthermore, we find that for a fixed \( q \), an increase in \( \alpha \) causes the function's minimum value to decrease, whereas an increase in \( \beta \) raises the minimum value. This may be attributed to the specific structure of the metric.

% \textbf{In short, $\alpha$ carries the legacy of dilaton physics rooted in string theory, whereas $\beta$ is a new ``mixing” parameter that allows us to interpolate between RN and dilaton black holes and to systematically study deviations from standard Kerr–Newman phenomenology.}

\section{Null geodesics and the Shadow}
In this section, we will constrain the parameter of $\alpha$ and $\beta$ by comparing the theoretically defined shadow radius with observational data from the EHT. We begin by considering a static, spherically symmetric spacetime as presented in Ref.\cite{177}
%first explore the behavior of the shadow radius theoretically, and then find the constraints to the EMS parameter using the data from EHT.  To begin with, consider a static, spherically symmetric spacetime given by  \cite{177}
\begin{equation}
 \begin{aligned}
d s^2=-A(r) d t^2+B(r) d r^2+C(r) \left(d \theta^2+ \sin ^2 \theta d \phi^2\right).
\end{aligned}   
\end{equation}
The functions \( A(r) \), \( B(r) \), and \( C(r) \) satisfy the asymptotic flatness conditions, with \( A(r) \) and \( B(r) \) $\rightarrow1$ at spatial infinity while \( C(r)\rightarrow r^2\). %Without loss of generality, we analyze the null geodesic in the equatorial plane only such that the polar angle is fixed to $\theta=\pi/2$.   Then, the Hamiltonian for light ray is given by  
For simplicity, we restrict our analysis to null geodesics in the equatorial plane by fixing the polar angle at $\theta=\pi/2$. Under this configuration, the photon's Hamiltonian takes the form
\begin{equation}
H=\frac{1}{2} g^{i k} p_i p_k=\frac{1}{2}\left(-\frac{p_t^2}{A(r)}+\frac{p_r^2}{B(r)}+\frac{p_\phi^2}{C(r)}\right) .  
\end{equation}
It is worth noting that since we consider EMS BH, the metric functions satisfy \( A(r) = f(r) \), \( B(r) = f(r)^{-1} \), and \( C(r) = r^2\left(1-\frac{b_2}{r}\right)^{\frac{2 \alpha^2}{1+\alpha^2}} \). Subsequently, the equations of motion for null particles are 
\begin{equation}
 \left\{\begin{split}
&\dot{x}^i=\frac{\partial H}{\partial p_i},\\
&\dot{p}_i=-\frac{\partial H}{\partial x^i}.\\
	\end{split}\right. \label{700}  
\end{equation}
%Here, $\dot{x}=dx/d \lambda$ and $\dot{p}$ represents the conjugate momenta. Eq.(\ref{700}) gives
Using Eq.\eqref{700}, we can derive
\begin{equation}
 \left\{\begin{split}
&\dot{t}=-\frac{p_t}{f(r)}, \quad \dot{\phi}=\frac{p_\phi}{C(r)}, \quad \dot{r}=-\frac{p_r}{f(r)^{-1}}, \\
&\dot{p}_t=0,\quad\dot{p}_\phi=0,\\
&\dot{p}_r=\frac{1}{2}\left(-\frac{p_t^2 f^{\prime}(r)}{f(r)^2}+\frac{p_r^2 f^{\prime}(r)^{-1}}{f(r)^{-2}}-\tilde{C} p_\phi^2\right).\\
	\end{split}\right. \label{801}
\end{equation}
where $\tilde{C}\equiv \frac{d}{dr}\frac{1}{C(r)}=-\frac{2\left(1-\frac{b_2}{r}\right)^{-\frac{2 \alpha^2}{1+\alpha^2}}}{r^3}-\frac{2 \alpha^2 b_2\left(1-\frac{b_2}{r}\right)^{-1-\frac{2 \alpha^2}{1+\alpha^2}}}{\left(1+\alpha^2\right) r^4}$. Setting $H=0$, we have 
\begin{equation}
-\frac{p_t^2}{f(r)}+\frac{p_r^2}{f(r)^{-1}}+\frac{p_\phi^2}{C(r)}=0, \label{3}
\end{equation}
and it now follows that 
\begin{equation}
\frac{dr}{d\phi}=\frac{\dot{r}}{\dot{\phi}}=\frac{C(r)}{f(r)^{-1}} \frac{p_{r}}{p_{\phi}}. \label{18}
\end{equation}
%Setting $p_{t}=-\omega_{0}$, and using $p_r$, we get the relation how $r$ changes with $\phi$:
Substituting the expressions for $p_r$ and $p_t=-\omega_0$, we finally obtain
\begin{equation}
    \frac{dr}{d\phi}=\pm \frac{r^2}{C(r)^{1/2} f(r)^{-1/2}}\sqrt{\frac{\omega^2_{0}}{p^2_{\phi}}h(r)^2-1},
\end{equation}
where $h(r)^2=\frac{C(r)}{f(r)}$ is defined.

For a circular photon orbit, the conditions \(\dot{r} = 0\) (vanishing radial velocity) and \(\ddot{r} = 0\) (zero radial acceleration) necessarily imply \(p_r = 0\), which reduces Eq.(\ref{3}) to the following form
% For a circular light orbit, the radial velocity and acceleration should be $\dot{r}=0$ and $\ddot{r}=0$ respectively, and hence, $p_{r}=0$. Eq.(\ref{3}) then becomes 
\begin{equation}
    0=-\frac{w^2_{0}}{f(r)}+\frac{p^2_{\phi}}{C(r)}.\label{4}
\end{equation}

Furthermore, Eq.(\ref{4}) can be expressed as 
\begin{equation}
    \dot{p}_{r}=\frac{\omega_{0}^2 f'(r)}{f(r)^2}+\tilde{C}p^2_{\phi}=0.\label{5}
\end{equation}

Using Eq.(\ref{4}) and (\ref{5}), we obtain 
\begin{equation}
    p^2_{\phi}=C(r)\frac{\omega^2_{0}}{f(r)}=-\frac{\omega_{0}^2 f'(r)}{\tilde{C} f(r)^2},\label{6}
\end{equation}
% and
% \begin{equation}
%     p^2_{\phi}=-\frac{\omega_{0}^2 f'(r)}{\tilde{C} f(r)^2}.\label{7}
% \end{equation}

Combining Eq.\eqref{5} and Eq.\eqref{6}, we give analytic expression for the photon sphere radius of the BH \cite{176,184} (See more details in Appendix A).
%The implication of subtracting Eqs. (\ref{6})  give the information on how to find the radius of the photon sphere \cite{176,184}

\begin{equation}
    0=\frac{d}{dr}h(r)^2. \label{8}
\end{equation}

Following Eq.(\ref{8}), the analytical form of the photon sphere radius $r_{\text{ph}}$ is so complicated. 
We plot the above equation numerically (See Fig.2), where it shows the values of $\alpha$, $\beta$ and $q$ for the photon sphere to exist. When $q=0.5$, as the parameter $\alpha$ increases, the photon sphere radius essentially increases. Then, for a given value of $\alpha$, we observe an decrease  in photon sphere radius as the $\beta$ increases.  The second picture is exactly opposite from the first picture.  When $\alpha=0$, as the parameter $q$ increases, the photon sphere radius essentially decreases. As $q$ approaches 0, the photon sphere radius asymptotically tends to 3 (Schwarzschild BH). Then, for a given value of $q$, we can observe an increase in photon sphere radius as the $\beta$ decreases. And when $\beta=0$,  as the parameter $q$ increases, the photon sphere radius essentially decreases. As $q$ approaches 0, the photon sphere radius asymptotically tends to 3. Then, for a given value of $q$, we can observe an increase in photon sphere radius as the $\alpha$ increases.

\begin{figure}[H]
	\begin{minipage}{0.5\textwidth}
    \centering
		\includegraphics[scale=0.9,angle=0]{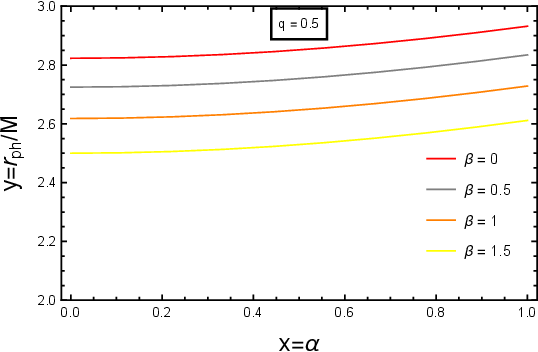}
	\end{minipage}
    \begin{minipage}{0.5\textwidth}
    \centering
		\includegraphics[scale=0.9,angle=0]{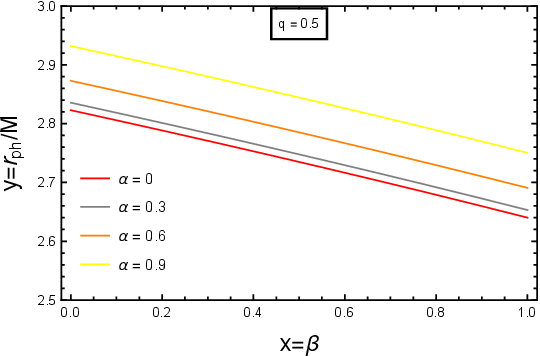}
	\end{minipage}	
    
    \begin{minipage}{0.5\textwidth}
		\includegraphics[scale=0.9,angle=0]{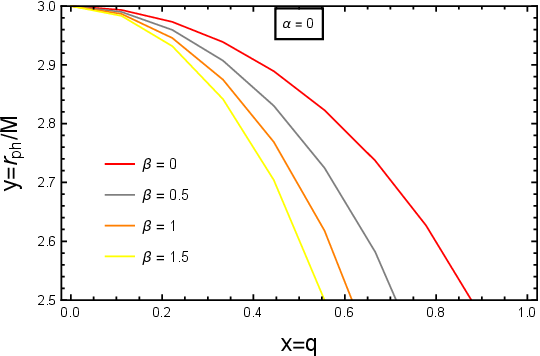}
	\end{minipage}
    \begin{minipage}{0.5\textwidth}
      \centering
		\includegraphics[scale=0.9,angle=0]{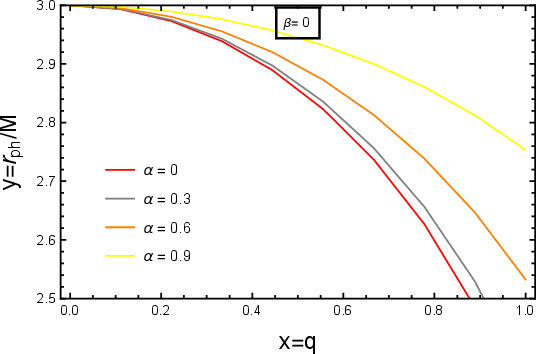}
	\end{minipage} \label{1000}
\caption{The above four pictures plot the photon sphere radius under the different parameters $\alpha$, $\beta$ and $q$.}
\end{figure}

% Let us now determine the behavior of the shadow radius. For constructing the shadow, we assume that a static observer at radius coordinate $r_{\text{o}}$ sends light rays into the past. As can be seen from  Fig.\ref{Fig.3}, the angle $a$ between such a light ray and the radial direction is given by
Subsequently, the definition of the shadow radius must be established. The shadow radius behavior can be investigated by considering light rays traced backward in time from a static observer located at radial coordinate \( r_O \). As shown in Fig.3, the angle \( a \) formed between such light rays and the radial direction is expressed as 
\begin{equation}
\cot a=\left.\frac{\sqrt{g_{r r}}}{\sqrt{g_{\phi \phi}}} \frac{d r}{d \phi}\right|_{r=r_{\mathrm{O}}}=\left.\frac{\sqrt{B(r)}}{\sqrt{C(r)}} \frac{d r}{d \phi}\right|_{r=r_{\mathrm{O}}}.
\end{equation}

With the help of Eq.(\ref{18}), we obtain \cite{176}
\begin{equation}
 \left\{\begin{split}
&\cot ^2 a=\frac{h\left(r_{\mathrm{O}}\right)^2}{h(R)^2}-1, \\
&\sin ^2 a=\frac{h(R)^2}{h\left(r_{\mathrm{O}}\right)^2}.\\
\end{split}\right. \label{8010}
\end{equation}
where $R$ means the minimum radius that the light ray approaches the center and then goes out (turning point of the trajectory).
% The boundary curve of the shadow corresponds to past-oriented light rays that asymptotically approach one of the unstable circular light orbits at radius $r_{\mathrm{ph}}$. Therefore we have to consider the limit $R \rightarrow r_{\mathrm{ph}}$ in Eq.(\ref{801}) for getting the angular radius $a_{\text {sh }}$ of the shadow \cite{184},
The shadow boundary is determined by past-oriented light rays asymptotically approaching the unstable circular photon orbits at radius \( r_{\mathrm{ph}} \). Consequently, the angular radius \( a_{\text{sh}} \) of the shadow is obtained by taking the limit \( R \rightarrow r_{\mathrm{ph}} \) in Eq.\eqref{8010} \cite{184}.
\begin{equation}
  \sin ^2 a_{\mathrm{sh}}=\frac{h\left(r_{\mathrm{ph}}\right)^2}{h\left(r_{\mathrm{O}}\right)^2}.\label{27} 
\end{equation}

Here $h(r)$ is well-defined, $h(r)^2=C(r)/A(r)$. Note that the critical value $b_{\text {c}}$ of the impact parameter is connected with $r_{\mathrm{ph}}$ by Ref.\cite{176}
\begin{equation}
  b_{\mathrm{c}}=h\left(r_{\mathrm{ph}}\right).  
\end{equation}

Therefore, the Eq.(\ref{27}) can also be expressed as
\begin{equation}
  \sin ^2 a_{\mathrm{sh}}=\frac{b_{\mathrm{c}}^2}{h\left(r_{\mathrm{O}}\right)^2}, \quad \text { or } \quad \sin ^2 a_{\mathrm{sh}}=\frac{b_{\mathrm{c}}^2 A\left(r_{\mathrm{O}}\right)}{C\left(r_{\mathrm{O}}\right)}  .
\end{equation}

For a general static spherically symmetric asymptotically flat spacetime, Ref.\cite{176} demonstrates that when the observer's radial coordinate \( r_O \) is sufficiently large, the BH shadow radius approaches \( b_c \). This radius is determined by the location of the photon sphere and the specific form of the metric function \( f(r) \). In our paper, we will obey this consideration.

\begin{figure}[H]
	\begin{minipage}{0.8\textwidth}
    \centering
		\includegraphics[scale=0.5,angle=0]{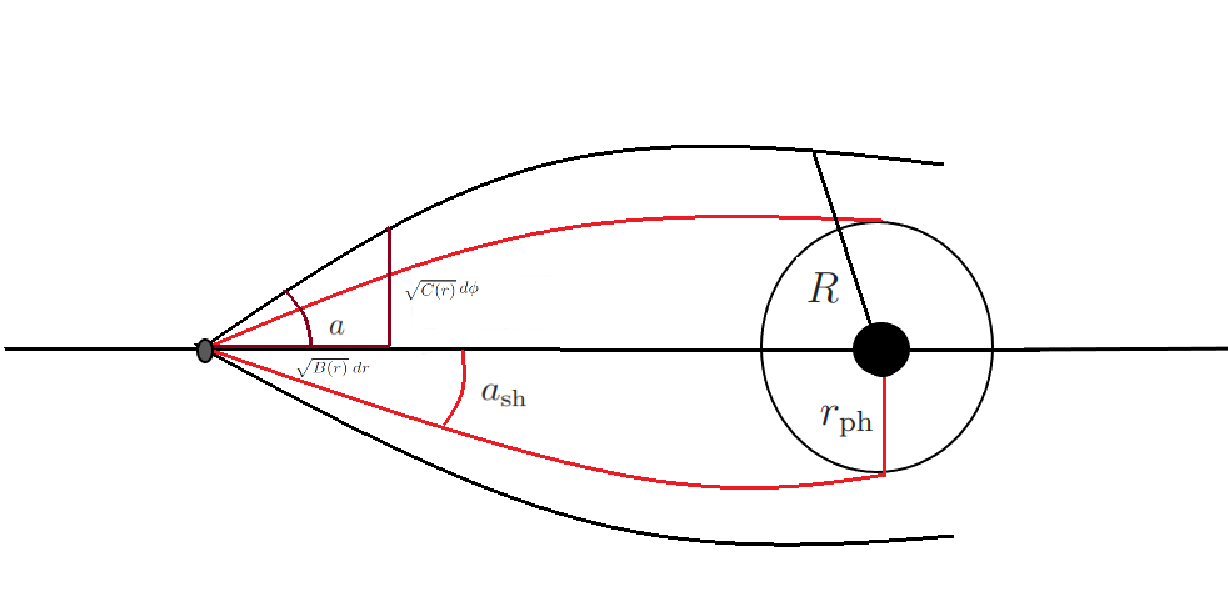}
	\end{minipage}
    \hspace{4cm}
\caption{Example of calculation of light ray emitted from the observer's position into the past under an angle $a$. The BH horizon and the photon sphere are shown; $r_{\mathrm{ph}}$ is the photon sphere radius. %The trajectory is calculated in the Asymptotically flat spacetime, $R$ denotes the radius coordinate at the point of closest approach. 
The picture is indicated in Ref.\cite{176}.}\label{Fig.3}
\end{figure}

\subsection{Constraints on parameters with the EHT observations of M87* and Sgr A*}
We will obtain constraints on the coupling parameters $\alpha$ and $\beta$ using EHT data for M87$^*$ and Sgr A$^*$, focusing on the non-rotating case. This simplification is justified because
\begin{itemize}
\item M87$^*$: General-relativistic magnetohydrodynamic simulations indicate that current EHT observations cannot reliably distinguish between Kerr and dilaton BHs based on shadow morphology alone~\cite{111}. The observed shadow size is consistent with $3\sqrt{3}(1 \pm 0.17)M$, regardless of spherical or axisymmetric models~\cite{112}.
\item Sgr A$^*$: The rotation parameter is sufficiently small to negligibly affect the shadow radius~\cite{110}.
\end{itemize}

Based on observational data, the shadow of M87* exhibits an angular diameter of \(42 \pm 3\ \mathrm{\rm \mu as}\), corresponding to a distance of \(16.8\) Mpc from Earth and a central BH mass of \((6.5 \pm 0.90) \times 10^9\ M_{\odot}\) \cite{113}. In parallel, the EHT's recent findings for Sgr A* report a shadow size of \(48.7 \pm 7\ \mathrm{\mu as}\), with the source located at a distance of \(8277 \pm 33\) pc and hosting a supermassive BH of mass \((4.3 \pm 0.013) \times 10^6\ M_{\odot}\) \cite{114,115}.

The shadow diameter in mass units can be calculated from the derived BH parameters using the following expression \cite{116}
% Now, once we have the above data about the BH, we can calculate the diameter of the shadow size in units of mass by using the following expression \cite{116},
\begin{equation}
    d_{\mathrm{sh}}=\frac{D \theta}{M}.
\end{equation}

Hence, the theoretical shadow diameter can be obtained via
\begin{equation}
 d_{\text {sh }}^{\text {theo }}=2 R_{\text {sh}}.  
\end{equation}
Therefore, by using the above expression, we get the diameter of the shadow of M87* and Sgr A* 
\begin{equation}
 \left\{\begin{split}
&d_{\mathrm{sh}}^{\mathrm{M} 87^*}=(11 \pm 1.5) M,\\
&d_{\mathrm{sh}}^{\mathrm{Sgr}.\mathrm{A}^*}=(9.5 \pm 1.4) M.\\
\end{split}\right.  
\end{equation}

Our constraints on the model parameters are shown in Fig.4. The corresponding numerical constraints are further presented in Tab.I (for M87*) and Tab.II(for Sgr A*).
\begin{figure}[H]
	\begin{minipage}{0.5\textwidth}
    \centering
		\includegraphics[scale=0.9,angle=0]{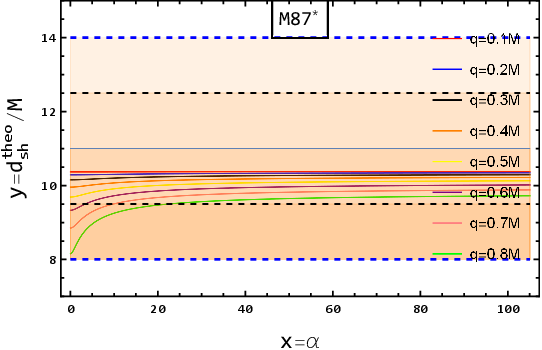}
	\end{minipage}
    \begin{minipage}{0.5\textwidth}
    \centering
 		\includegraphics[scale=0.9,angle=0]{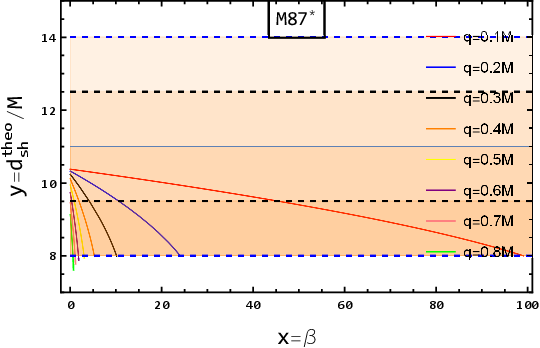}
	\end{minipage}	
    
	\begin{minipage}{0.5\textwidth}
    \centering
		\includegraphics[scale=0.9,angle=0]{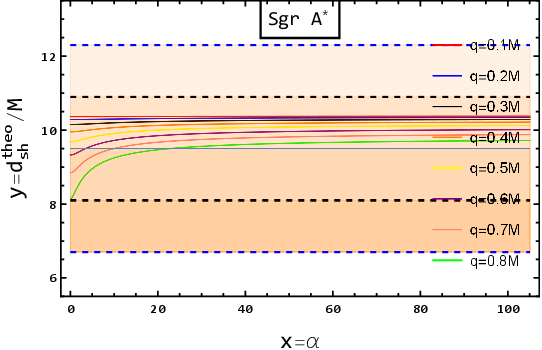}
	\end{minipage}
    \begin{minipage}{0.5\textwidth}
    \centering
		\includegraphics[scale=0.9,angle=0]{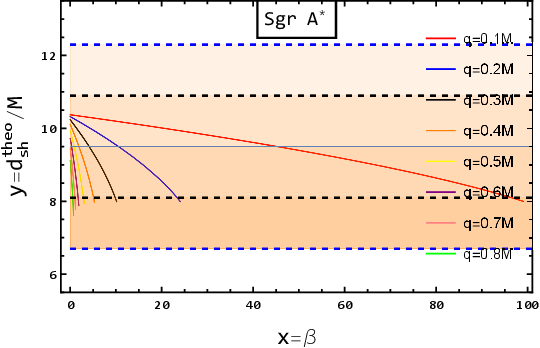}
	\end{minipage}	
\caption{These plots are showing the constraints for  different coupling parameters $\alpha$ and $\beta$.}\label{Fig.4}
\end{figure}

\begin{center}

\begin{tabular}{|c|ccccc|c|cccc|}
\hline
\hline M87*$/\alpha$ & $2 \sigma$ & $2 \sigma$   & $1 \sigma$ &   $1 \sigma$&&  M87*/$\beta$& $2 \sigma$ & $2 \sigma$   & $1 \sigma$ &   $1 \sigma$ \\
\hline charge $Q$ & upper  & lower & upper & lower& &charge $Q$& upper  & lower & upper & lower                                   \\
\hline 0.1 M & - & - & -&-     &&0.1M & 99.00  & - & 45.32&-                        \\
\hline 0.2 M & - & - & -&-     &&0.2M &  24.00 & - & 10.58&-                           \\
\hline 0.3 M & - & - & -&-    & &0.3M &  10.11 & - & 4.149&-                          \\
\hline 0.4 M & - & - & -&-    & &0.4M &  5.255 & - & 1.899&-                          \\
\hline 0.5 M & - & - & -&-    & &0.5M & 3.008 & - & 0.8592&-                            \\
\hline 0.6 M & - & - & -&3.441& &0.6M &  1.790 & - & 0.2960&-                         \\
\hline 0.7 M & - & - & -&9.889& &0.7M &  1.058 & - & -&-                         \\
\hline 0.8 M & - & - & -&22.59& &0.8M &  0.5866 & - & -&-                           \\
\hline
\end{tabular}

 \par\vspace{8pt}
TABLE I. The table lists the values of $\alpha$ and $\beta$ at the $1\sigma$ and $2\sigma$ confidence levels for M87*.
\end{center}

\begin{center}

\begin{tabular}{|c|ccccc|c|cccc|}
\hline
\hline Sgr A*$/\alpha$ & $2 \sigma$ & $2 \sigma$   & $1 \sigma$ &   $1 \sigma$ & & Sgr A*/$\beta$& $2 \sigma$ & $2 \sigma$   & $1 \sigma$ &   $1 \sigma$ \\
\hline charge $Q$ & upper  & lower & upper & lower &&charge $Q$& upper  & lower & upper & lower                                   \\
\hline 0.1 M & - & - & -&-   &  &0.1M& - & - &96.42  &-                       \\
\hline 0.2 M & - & - & -&-   &  &0.2M & - & - &23.36  &-                           \\
\hline 0.3 M & - & - & -&-   &  &0.3M & - & - &9.827   &-                         \\
\hline 0.4 M & - & - & -&-   &  &0.4M & - & - &5.094  &-                          \\
\hline 0.5 M & - & - & -&-   &  &0.5M & - & - &2.905   &-                            \\
\hline 0.6 M & - & - & -&-& &0.6M & - & - &1.718  &-                          \\
\hline 0.7 M & - & - & -&-& &0.7M & - & - &1.005   &-                       \\
\hline 0.8 M & - & - & -&- &&0.8M & - & - &0.5457   &-                       \\
\hline
\end{tabular}

 \par\vspace{8pt}
TABLE II. The table lists the values of $\alpha$ and $\beta$ at the $1\sigma$ and $2\sigma$ confidence levels for Sgr A*.
\end{center}

The variation of the diameter of the shadow image with coupling parameter $\alpha$ and $\beta$ for M87* and for Sgr A* is shown in Fig.\ref{Fig.4}, showing uncertainties at $1\sigma$ and $2\sigma$ levels. The numerical values for the upper or lower bounds in $\beta$ and $\alpha $ is found in Tab.I and Tab.II.
However, it turns out that the data for M87* gives a better constraint for the coupling parameter $\beta$. It would mean that there is a certain value for the coupling parameter that gives the observed value of M87* shadow. For the case of Sgr A*, we find that the parameter \( \alpha \) remains unconstrained within the \( 1\sigma \) and \( 2\sigma \) confidence intervals, whereas the parameter \( \beta \) is well constrained within the \( 1\sigma \) interval.

\subsection{Light bending: direct emission, lensing ring and photon ring}
To systematically investigate the observational characteristics of EMS BH, we examine their photon ring and lensing ring structures under the assumption of an optically thick accretion disk \cite{177,178,179}. Our analysis begins with studying null geodesics in the EMS BH. 

For computational convenience, we employ the coordinate transformation $u \equiv 1/r$, which yields the compact orbital equation
\begin{equation}
    \left(\frac{du}{d\phi}\right)^2 = G(u), \label{311}
\end{equation}
where the function $G(u)$ takes the explicit form
\begin{equation}
   \sqrt{G(u)} = \frac{1}{C(1/u)^{1/2} f(1/u)^{-1/2}u^2}\sqrt{\frac{\omega^2_{0}}{p^2_{\phi}}h(1/u)^2 - 1}. \label{802}
\end{equation}

Similarly, we restrict our analysis to equatorial motion. The system's Lagrangian formalism provides two conserved quantities, as derived in Eq.(\ref{801}). The critical impact parameter $b_c$ separates distinct null geodesic behaviors
\begin{itemize}
    \item Case 1 ($b > b_c$): Photons from asymptotic infinity are gravitationally scattered, reaching a minimum coordinate distance $r_{\text{min}}$ before returning to infinity.
    \item Case 2 ($b < b_c$): Photons become captured by the BH, demonstrating the existence of a photon capture cross-section.
    \item Case 3 ($b = b_c$): Photons spiral asymptotically toward the photon sphere at $r_{\text{ph}}$, defining the boundary of BH shadow.
\end{itemize}

%The impact parameter is defined as \( b \equiv \frac{p_t}{p_{\phi}} \). When \( b > b_c \), a light ray originating from infinity approaches the black hole, reaches a closest approach, and then escapes back to infinity. For \( b < b_c \), the light ray inevitably falls into the black hole. In the critical case \( b = b_c \), the light ray asymptotically orbits the black hole at the photon sphere radius \( r_{\mathrm{ph}} \).  

For \( b > b_c \), the turning point of the trajectory corresponds to the smallest positive real root of \( G(u) = 0 \), denoted as \( u_{\rm m} =1/r_{\rm min} \) with $r_{\rm min}$ being the light ray's radial minimal distance from its trajectory to the BH. Hence, the total change in the azimuthal angle \( \phi \) outside the horizon of the  trajectory with a given impact parameter \( b \) can be expressed as \cite{178,179}
\begin{equation}
   \phi=2 \int_0^{u_{\rm m}} \frac{d u}{\sqrt{G(u)}}, \quad b>b_{c}.
\end{equation}

For $b<b_{c}$,  the total change of angle $\phi$ is given by
\begin{equation}
  \phi=\int_0^{u_0} \frac{d u}{\sqrt{G(u)}}, \quad b<b_{c}, 
\end{equation}
where $u_0\equiv1 /r_+$.

\begin{figure}[H]
\centering
	\begin{minipage}{0.3\textwidth}
    \centering
		\includegraphics[scale=0.5,angle=0]{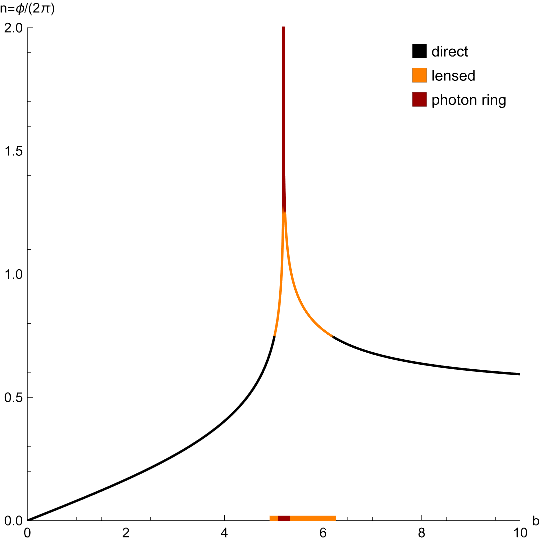}
	\end{minipage}
    \begin{minipage}{0.3\textwidth}
    \centering
 		\includegraphics[scale=0.5,angle=0]{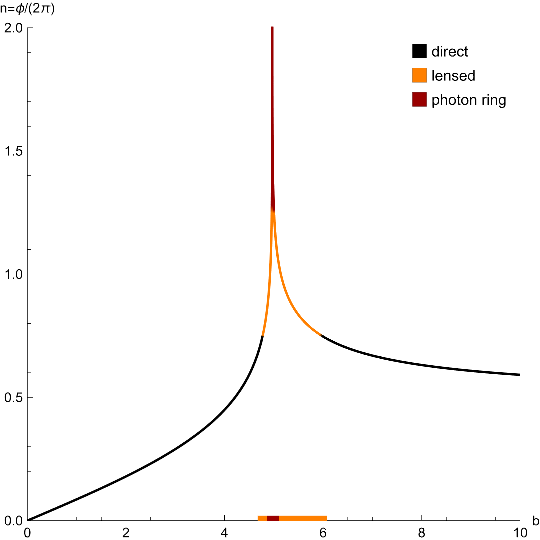}
	\end{minipage}	
    \begin{minipage}{0.3\textwidth}
    \centering
		\includegraphics[scale=0.5,angle=0]{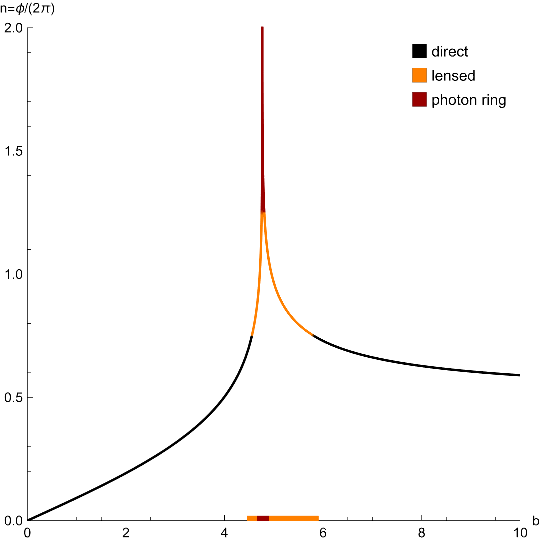}
	\end{minipage}
  \begin{minipage}{0.3\textwidth}
    \centering
		\includegraphics[scale=0.5,angle=0]{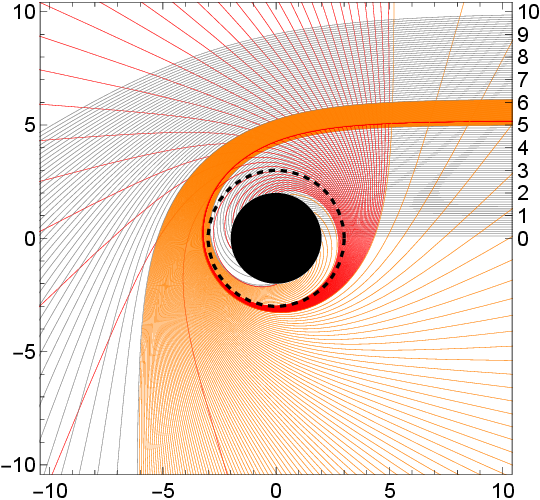}
	\end{minipage}
    \begin{minipage}{0.3\textwidth}
    \centering
		\includegraphics[scale=0.5,angle=0]{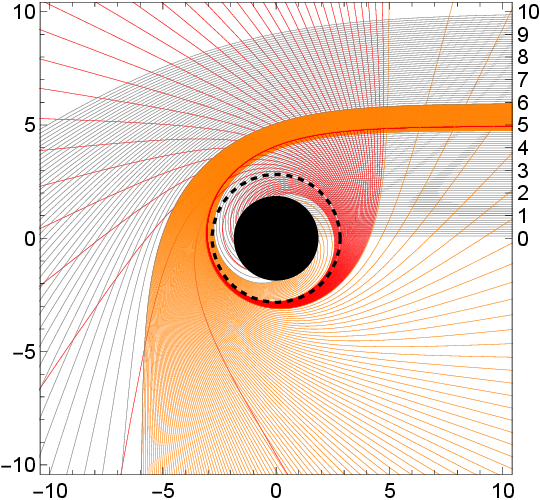}
	\end{minipage}
    \begin{minipage}{0.3\textwidth}
    \centering
		\includegraphics[scale=0.5,angle=0]{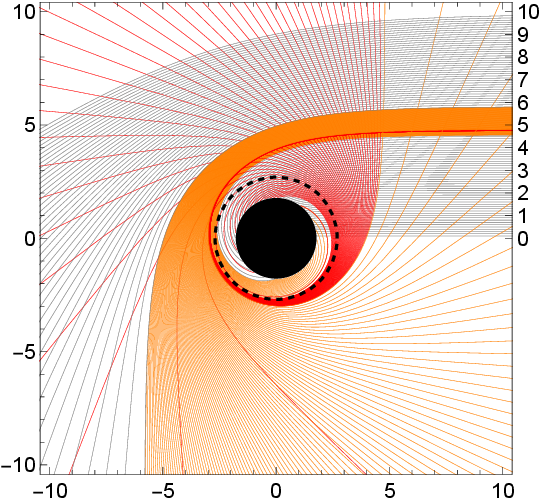}
	\end{minipage}	
\caption{The behavior of photon trajectories around the Schwarzschild, RN, and EMS BHs as a function of the impact parameter $b$.  In the upper panel, we present the total number of orbits, defined as  $n = \phi / 2\pi $. The trajectories are categorized based on $n $, where direct emission $ n < 3/4 $ is shown in black, lensed trajectories $3/4 < n < 5/4 $ in yellow, and photon ring trajectories $n > 5/4$ in red.  The lower panel displays selected photon trajectories in Euclidean polar coordinates $ (r, \phi) $. The spacing in the impact parameter is set to $ 1/10 $, $ 1/100 $, and $ 1/1000 $ for the direct, lensed, and photon ring trajectories, respectively. The BH is represented as a solid disk, while the dashed black circle in the ray-tracing diagram marks the photon orbit. For the three cases studied, we set $ \alpha = 0, \beta = 0, q = 0 $ for the Schwarzschild BH (first column), $ \alpha = 0, \beta = 0, q = 0.5 $ for the RN BH (second column), and $ \alpha = 0.5, \beta = 0.8, q = 0.5 $ for the EMS BH (third column). } \label{Fig.6}
\end{figure}

%To analyze the observational signatures of radiation emitted from the vicinity of a BH accretion disk. 
Following the trajectory classification established in Refs.~\cite{120,178} where photon orbits are categorized into direct, lensed, and photon ring components, we quantify the orbital winding through the turn number $n(b) \equiv \phi/2\pi$. This quantity exhibits discrete quantization as demonstrated in Refs.~\cite{178,179}, with the functional dependence on impact parameter $b$ given by
\begin{equation}
n(b) = \frac{2m - 1}{4}, \quad m \in \mathbb{Z}^+.
\end{equation}
The critical impact parameters $b_m^\pm$ demarcate these orbital families, where $b_m^-$ (satisfying $b_m^- < b_c$) and $b_m^+$ (satisfying $b_m^+ > b_c$) correspond respectively to the minimal and maximal solutions for each $m$-th order orbit, with $b_c$ denoting the photon sphere critical parameter.
%The Ref. \cite{120,178} divide trajectories into direct, lensed and photon rings. Now we can  define the total number of orbits $n=\frac{\phi}{2 \pi}$, which is obviously a function of impact parameter $b$, satisfying  \cite{178,179}
%\begin{equation}
%  n(b)=\frac{2 m-1}{4}, \quad m=1,2,3, \cdots  
%\end{equation}
%by $b_m^{ \pm}$ with $b_m^{-}<b_{c}$ and $b_m^{+}>b_{c}$ being the minimum and the maximum solutions, respectively. 
Then the rays can be classified  as follows
\begin{itemize}
    \item Direct: $\frac{1}{4}<n<\frac{3}{4} \Rightarrow b \in\left(b_1^{-}, b_2^{-}\right) \cup\left(b_2^{+}, \infty\right)$
    \item Lensed: $\frac{3}{4}<n<\frac{5}{4} \Rightarrow b \in\left(b_2^{-}, b_3^{-}\right) \cup\left(b_3^{+}, b_2^{+}\right)$
    \item Photon ring: $n>\frac{5}{4} \Rightarrow b \in\left(b_3^{-}, b_3^{+}\right)$
\end{itemize}

The behaviors of photons in  different BHs are plotted in Fig.5. The up panel of Fig.5 depicts the total number of orbits. And the down panel depicts the trajectories of light rays surrounding the BHs, such as "direct" will intersect the equatorial plane only once.

%The physical picture of this classification is clear from the trajectory plots in Fig.5. Assuming light rays emit from north pole direction (far right of the trajectory plots), trajectories whose number of orbits $1/4<n<3/4$ will intersect the equatorial plane only once. Trajectories whose number of orbits $3/4<n<5/4$ will intersect the equatorial plane twice. Trajectories whose number of orbits $n>5/4$ will intersect the equatorial plane at least 3 times.

\begin{center}
   \begin{tabular}{llll}
\hline \hline Parameter & $\alpha=0,\beta=0,q=0$ & $\alpha=0,\beta=0,q=0.5$ & $\alpha=0.5,\beta=0.8,q=0.5$ \\
\hline Direct Emission & $b<5.01514$ & $b<4.77294$ & $b<4.55452$ \\

              $n<3 / 4$ & $b>6.16757$ & $b>5.97448$ & $b>5.81006$ \\
\\
Lensing Ring & $5.01514<b<5.18781$ & $4.77294<b<4.95793$  &$4.55452<b<4.75392$ \\
$3 / 4<n<5 / 4$ & $5.22794<b<6.16757$ &$5.0039<b<5.97448$&$4.80685<b<5.81006$  \\
\\
Photon Ring \\
$n>5 / 4$              & $5.18781<b<5.22794$ & $4.95793<b< 5.0039$   
                                                    &$4.75392<b<4.80685$ \\
\hline
\end{tabular}

\par\vspace{8pt}

TABLE III. The region of direct emission, lensing ring, and photon ring for the different parameter $\alpha,\beta,q$. 
\end{center}

Table III illustrates the variation of the BH shadow with increasing parameters \( \alpha \), \( \beta \), and \( q \). From the table and corresponding figures, it is observed that the range of the lensing and photon rings expands progressively from the Schwarzschild BH to the RN BH and further to the EMS BH. This indicates that, from left to right, the contribution to the brightness of the lensing and photon rings increases accordingly.  Moreover, when the impact parameter approaches the critical value \(b\to b_c\), the photon orbit exhibits a sharp peak in the \( (b, \phi) \) plane. Beyond this regime, as \( b \) further increases, the photon trajectories correspond predominantly to direct emission across all cases considered.

\subsection{Transfer functions and observed specific intensities}
%We calculate the emitted intensity from the EMS black hole, modeling the disk radiation as isotropic in the static observer's reference frame. According to Liouville's theorem, the quantity \( I_\nu^{\mathrm{em}} / \nu_e^3 \) remains conserved along the trajectory of the light ray.
In this subsection, we will derive the specific intensity $I_\nu^{\mathrm{em}}$ radiated by the accretion disk, assuming local isotropy of radiation in the rest frame of static observers. %This treatment is fundamentally constrained by Liouville's theorem of relativistic radiative transfer, which dictates the invariant nature of the specific intensity-to-cubic-frequency ratio $I_\nu^{\mathrm{em}}/\nu_e^3$ along null geodesics in the phase space of photon propagation.
The observed intensity can be expressed as \cite{177,178,179}
\begin{equation}
  I_{\nu^{\prime}}^{\mathrm{obs}}=g^3 I_\nu^{\mathrm{em}},
\end{equation}
where $g=\sqrt{f(r)}$ and $I_{\nu^{\prime}}^{o b s}$ are the observed intensity at the frequency $\nu^{\prime}$. By integrating over all frequencies, we obtain the total emitted intensity as \( I^{\mathrm{em}} = \int I_\nu^{\mathrm{em}} d\nu \). Thus, the corresponding observed frequency can be written as
\begin{equation}
 I^{\mathrm{obs}}=g^4 I^{\mathrm{em}}.   
\end{equation}
$I^{\mathrm{em}}$  denotes the integrated specific intensity radiating from the entire accretion disk. Hence, the total received intensity is
\begin{equation}
  I^{\mathrm{obs}}(b)=\left.\sum_n g^4 I^{\mathrm{em}}(r)\right|_{r=r_m(b)},  
\end{equation}
where $r_m(b)$ is transfer function (It represents the $m$ intersection points with the equatorial plane). The transfer function fundamentally connects radial coordinates with photon impact parameters, we emphasize that our analysis operates under two key idealizations: (1) no light absorption or reflection by the disk, and (2) no environmental intensity attenuation.

Therefore, %We denote the solution of the orbit equation by \( u(\phi, b) \) and focus on 
the first three transfer functions, which can be obtained by  \cite{177,178,179}
\begin{equation}
\left\{\begin{split}
&r_1(b)=\frac{1}{u\left(\frac{\pi}{2}, b\right)}, & b \in\left(b_1^{-}, \infty\right),\\
&r_2(b)=\frac{1}{u\left(\frac{3 \pi}{2}, b\right)}, & b \in\left(b_2^{-}, b_2^{+}\right),\\
&r_3(b)=\frac{1}{u\left(\frac{5 \pi}{2}, b\right)}, & b \in\left(b_3^{-}, b_3^{+}\right).
\end{split}\right.  
\end{equation}

As established in Ref.\cite{120}, three distinct observational features emerge:(1) The first transfer function corresponds to the direct disk image;(2) The second forms a significantly demagnified lensing ring from the disk's far side;(3) The third creates the extremely faint photon ring from the near side. The corresponding graphical representation is presented in the figure below.
\begin{figure}[H]
\centering
	\begin{minipage}{0.3\textwidth}
    \centering
		\includegraphics[scale=0.55,angle=0]{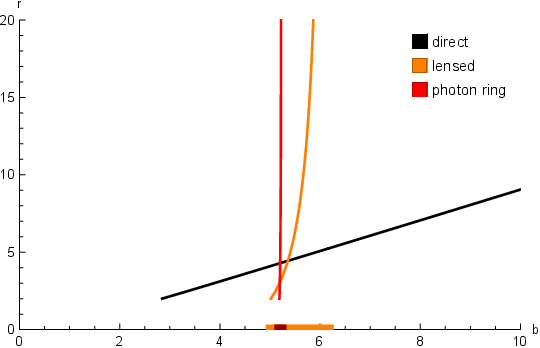}
	\end{minipage}
    \begin{minipage}{0.3\textwidth}
    \centering
 		\includegraphics[scale=0.55,angle=0]{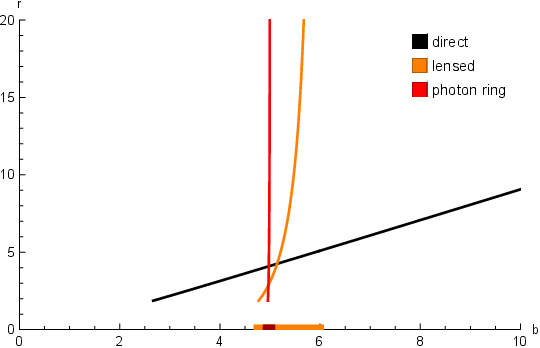}
	\end{minipage}	
    \begin{minipage}{0.3\textwidth}
    \centering
		\includegraphics[scale=0.55,angle=0]{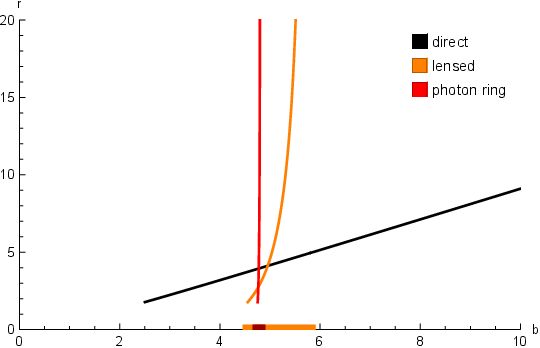}
	\end{minipage}
\caption{The first three transfer functions in BHs for different values of $\alpha, \beta,q$. The above figure, from left to right, respectively
  Schwarzschild BH, RN BH and EMS BH. Here, the y-axis is the transfer function $r_m(b)$ and the x-axis is the impact parameter b.} 
\end{figure}
% In Fig.6 , the black dots \((m=1)\) represent the transfer function for direct emission and exhibit an almost constant slope, indicating a redshifted source profile. The yellow dots \((m=2)\) correspond to the lensing ring, where the slope remains small near \(b_c\) but increases rapidly with \(b\), leading to significant demagnification of the image from the far side of the accretion disk. The red dots \((m=3)\) represent the photon ring, where the slope approaches infinity, implying extreme demagnification of the disk's near-side image. Consequently, the dominant contribution to the observed flux arises from direct emission, while higher-order images \((m>3)\) contribute negligibly and can be disregarded.
As shown in Fig.6, the black points $(m = 1)$ correspond to the direct emission transfer function and exhibit an approximately constant slope, reflecting the redshifted emission profile from the disk. The yellow points $(m = 2)$, associated with the lensing ring, display a rapidly increasing slope with growing impact parameter $b$, indicating that the far-side image is significantly demagnified. The red points $(m = 3)$ denote the photon ring, where the slope becomes nearly infinite, resulting in extreme demagnification of the near-side image. Thus, the observed flux is primarily governed by the direct emission, with higher-order contributions $(m > 3)$ being negligible.

All three BHs exhibit these characteristics. The critical impact parameters are \( b_c = 5.19615 \) for the Schwarzschild BH, \( b_c = 4.96791 \) for the RN BH, and \( b_c = 4.76598 \) for the EMS BH. This result indicates that the Schwarzschild BH has the largest critical impact parameter, corresponding to the largest shadow radius, while an increase in the model parameters leads to a reduction in the BH’s shadow size.

Additionally, we not only learn the aforementioned characteristics but can also quantify both the morphology (width) and intensity changes of the three ring types. First, we discuss how scalar couplings affect the photon ring. As shown in Table III, we have following calculations. In Schwarzschild BH ($q=0,\alpha=0,\beta=0$), we have $\Delta b_{\text{ph}} = b_3^+ - b_3^- = 5.22794 - 5.18781 = 0.04013$. In EMS BH ($\alpha=0.5,\beta=0.8,q=0.5$), we have $\Delta b_{\text{ph}} = b_3^+ - b_3^- = 4.80685 - 4.75392 = 0.05293$. The EMS BH shows a net increase of $32\%$ in width. A possible explanation is: the scalar field flattens the effective potential near  $r_{\text{ph}} 
$ (due to $C(r) = r^2(1-b_2/r)^{\frac{2\alpha^2}{1+\alpha^2}}$), allowing more near-critical orbits. 

For lensing ring adjustment, we now properly treat the two disjoint intervals ($b \in (b_2^-, b_3^-)$ and $b \in (b_3^+, b_2^+)$). In Schwarzschild BH ($q=0,\alpha=0,\beta=0$), we have $\Delta b_{\text{lens1}} = b_3^- - b_2^- = 5.18781 - 5.01514 = 0.17267$, $\Delta b_{\text{lens2}} = b_2^+ - b_3^+ = 6.16757 - 5.22794 = 0.93963$. Total $\Delta b_{\text{lens}} = 1.1123$. In EMS BH ($\alpha=0.5,\beta=0.8,q=0.5$), we have  $\Delta b_{\text{lens1}} = b_3^- - b_2^- = 4.75392 - 4.55452 = 0.1994$, $\Delta b_{\text{lens2}} = b_2^+ - b_3^+ = 5.81006 - 4.80685 = 1.00321$. Total $\Delta b_{\text{lens}} = 1.20261$.  Compared to a Schwarzschild BH, the EMS BH shows a net increase of $8.1\%$.  While the total lensing ring width increases modestly, the inner sub-interval ($\Delta b_{\text{lens1}}$) shows a $15.5\%$ increase due to stronger light bending near the photon sphere. 

Finally we consider  the direct emission, increasing $\alpha$ leads to a larger $b_{c}$, whereas increasing $\beta$ decreases $b_{c}$. From Table III, the boundaries of the direct emission region exhibit a systematic shift along the impact-parameter axis.

%\textbf{We noted that while the slopes of the transfer function $r_3(b)$ for both Schwarzschild and EMS black holes approach infinity at the critical point, the EMS black hole's curve is overall steeper in the physical region where the rings are formed. This implies that the EMS black hole exhibits a stronger demagnification effect on higher-order images. Consequently, photons from a larger area on the accretion disk are compressed into a smaller range of impact parameters on the observer's screen, which significantly reduces the brightness of the photon ring and lensing rings. Thus, from Figs. 7–9, EMS rings appear  dimmer than Schwarzschild across models, with suppression dominating any flux increase from wider rings.}

\subsection{Observational features of direct emission, photon and lensing rings}
As discussed in the previous section, for an observer at infinity, the specific intensity depends solely on the radial coordinate $r $. In our analysis, we consider the following  three toy models for the emissivity profile $ I^{\mathrm{em}} $\cite{177,178,179},
%With the transfer function determined, we now consider a specific emission profile. As the first model, we examine a scenario in which the emission is sharply concentrated at the innermost stable circular orbit (ISCO) given by
\begin{equation}
I_{1}^{\mathrm{em}}(r)= \begin{cases}\left(\frac{1}{r-\left(r_{\mathrm{isco}}-1\right)}\right)^2, & r > r_{\mathrm{isco}} \\ 0, & r \leq r_{\mathrm{isco}}\end{cases}
\end{equation}
%Second model, we consider the emission is sharply peaked at the photon sphere, and it ends abruptly at $r=r_{\text{ph}}$ while quickly decaying to zero, such as
\begin{equation}
I_{2}^{\mathrm{em}}(r)= \begin{cases}\left(\frac{1}{r-\left(r_{\mathrm{ph}}-1\right)}\right)^3, & r > r_{\mathrm{ph}} \\ 0, & r \leq r_{\mathrm{ph}}\end{cases}
\end{equation}
%Third model, we consider an emission decaying gradually from the horizon to the ISCO, such as 
\begin{equation}
I_{3}^{\mathrm{em}}(r)= \begin{cases}\frac{1-\arctan \left(r-\left(r_{\text {isco }}-1\right)\right)}{1-\arctan \left(r_{\mathrm{ph}}\right)}, & r >r_{\mathrm{+}} \\ 0, & r \leq r_{\mathrm{+}}\end{cases}
\end{equation}
%where \( r_{\text{isco}} \) denotes the innermost stable circular orbit.
where $r_{\text {isco}}, r_{\text {ph }}$ and $r_{\mathrm{+}}$ is the innermost stable orbit, photon sphere and event horizon respectively. These three models each exhibit distinct characteristics. %The second model shows a rapid decay behavior, whereas the third model decays much more slowly. In terms of emission origin, the third model features radiation emerging directly from the event horizon, the second model begins at the photon sphere, and in the first model, the emission is confined near the innermost stable circular orbit (ISCO).
The following figures present the results for three distinct types of BHs.

\begin{figure}[H]
\centering
	\begin{minipage}{0.3\textwidth}
    \centering
		\includegraphics[scale=0.55,angle=0]{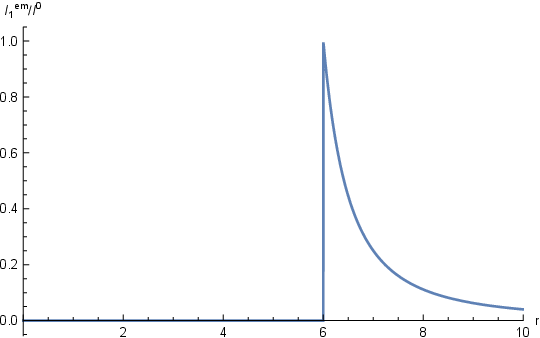}
	\end{minipage}
    \hspace{0.5cm}
    \begin{minipage}{0.3\textwidth}
    \centering
 		\includegraphics[scale=0.55,angle=0]{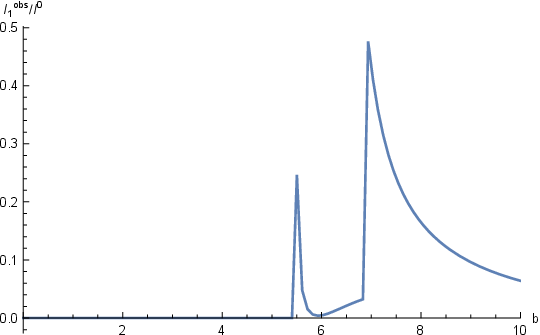}
	\end{minipage}	
      \hspace{0.5cm}
    \begin{minipage}{0.3\textwidth}
    \centering
     \vspace{-0.5cm}
		\includegraphics[scale=0.55,angle=0]{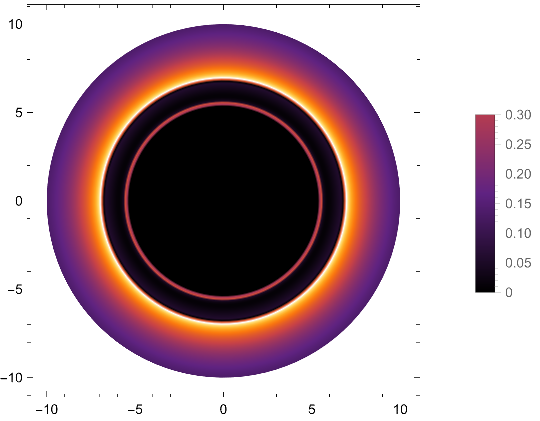}
	\end{minipage}

\vspace{2em}

\begin{minipage}{0.3\textwidth}
    \centering
		\includegraphics[scale=0.55,angle=0]{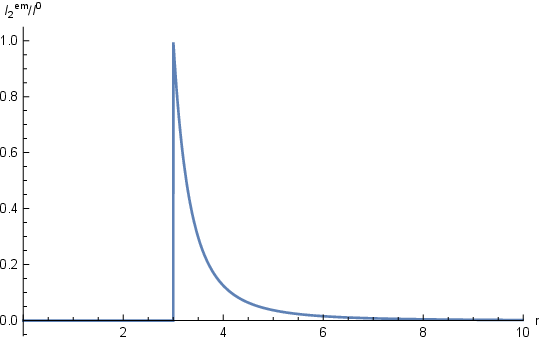}
	\end{minipage}
    \hspace{0.5cm}
    \begin{minipage}{0.3\textwidth}
    \centering
 		\includegraphics[scale=0.55,angle=0]{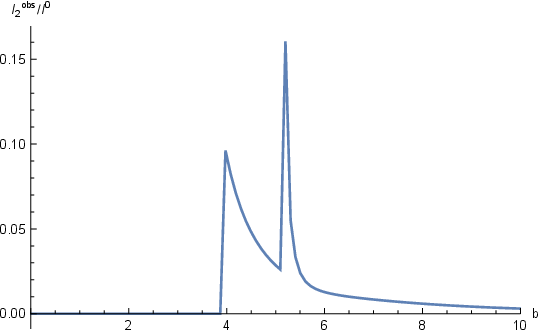}
	\end{minipage}	
      \hspace{0.5cm}
    \begin{minipage}{0.3\textwidth}
    \centering
    \vspace{-0.5cm}
		\includegraphics[scale=0.55,angle=0]{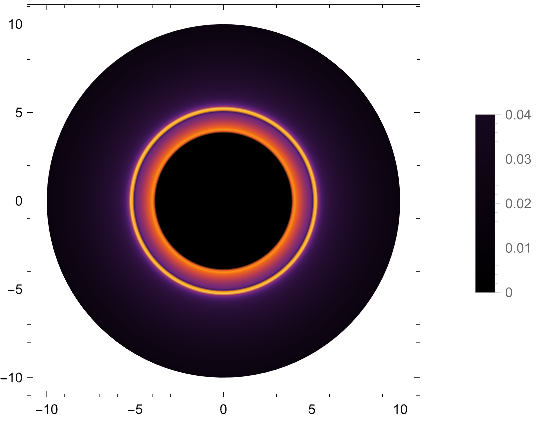}
	\end{minipage}

\vspace{2em}

\begin{minipage}{0.3\textwidth}
    \centering
		\includegraphics[scale=0.55,angle=0]{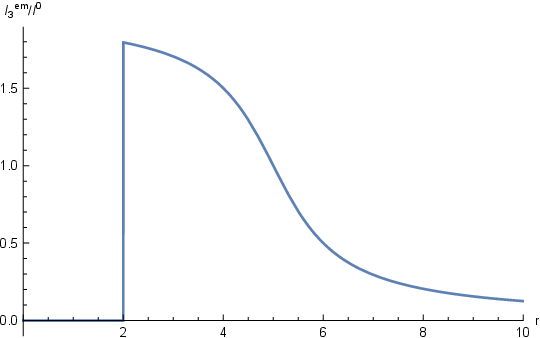}
	\end{minipage}
    \hspace{0.5cm}
    \begin{minipage}{0.3\textwidth}
    \centering
 		\includegraphics[scale=0.55,angle=0]{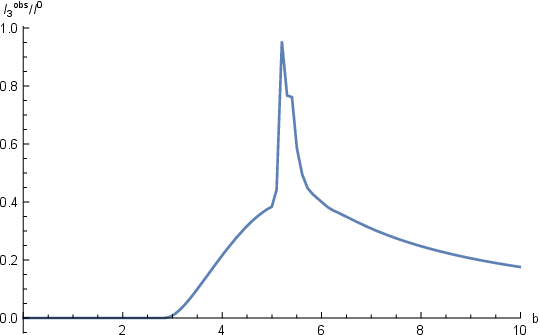}
	\end{minipage}
      \hspace{0.5cm}
    \begin{minipage}{0.3\textwidth}
    \centering
    \vspace{-0.5cm}
		\includegraphics[scale=0.55,angle=0]{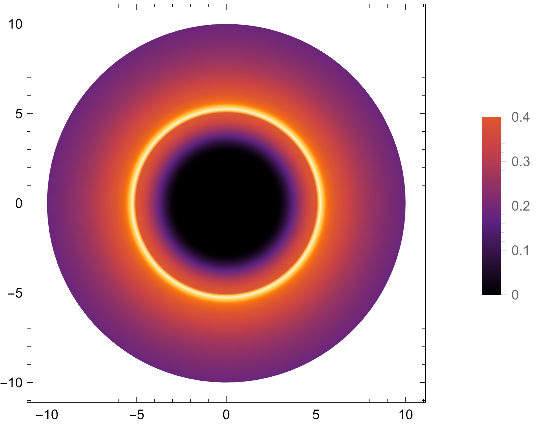}
	\end{minipage}
\caption{ The observational appearance of the thin disk with different emission profiles for \( \alpha = 0, \beta = 0, q = 0 \) is shown from a face-on perspective. The first row corresponds to the emission profile intensity as described by model 1, the second row represents model 2, and the third row corresponds to model 3, as outlined in Section III.C. In the plots, the emitted and observed intensities, \( I^{\mathrm{em}} \) and \( I^{\mathrm{obs}} \), are normalized to the maximum value \( I^0 \) of the emitted intensity outside the event horizon. } 
\end{figure}

\begin{figure}[H]
\centering
	\begin{minipage}{0.3\textwidth}
    \centering
		\includegraphics[scale=0.55,angle=0]{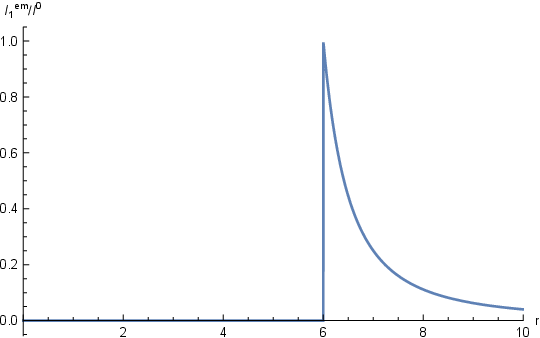}
	\end{minipage}
    \hspace{0.5cm}
    \begin{minipage}{0.3\textwidth}
    \centering
 		\includegraphics[scale=0.55,angle=0]{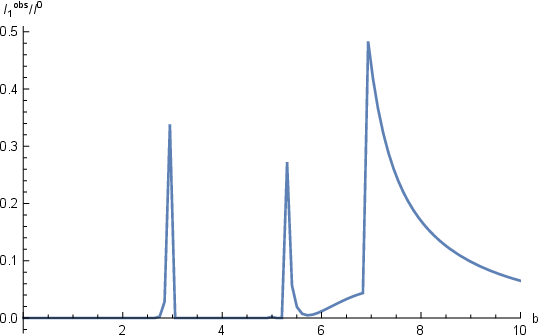}
	\end{minipage}	
      \hspace{0.5cm}
    \begin{minipage}{0.3\textwidth}
    \centering
     \vspace{-0.5cm}
		\includegraphics[scale=0.55,angle=0]{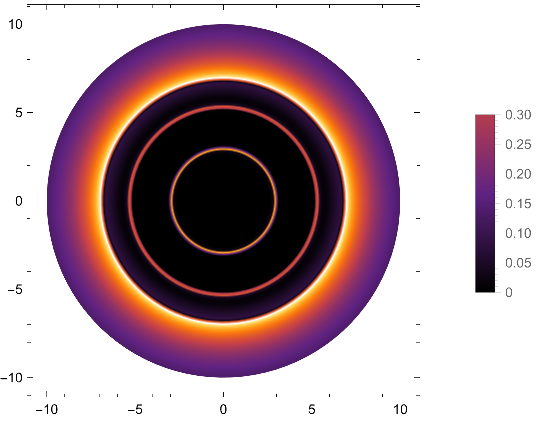}
	\end{minipage}

\vspace{2em}

\begin{minipage}{0.3\textwidth}
    \centering
		\includegraphics[scale=0.55,angle=0]{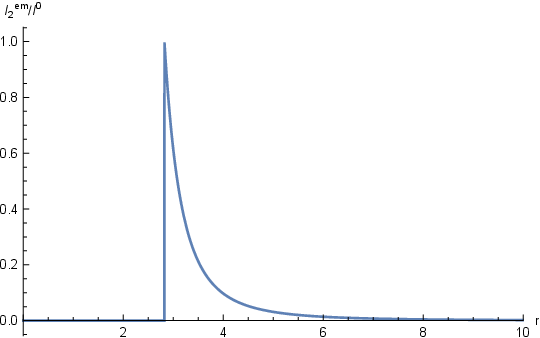}
	\end{minipage}
    \hspace{0.5cm}
    \begin{minipage}{0.3\textwidth}
    \centering
 		\includegraphics[scale=0.55,angle=0]{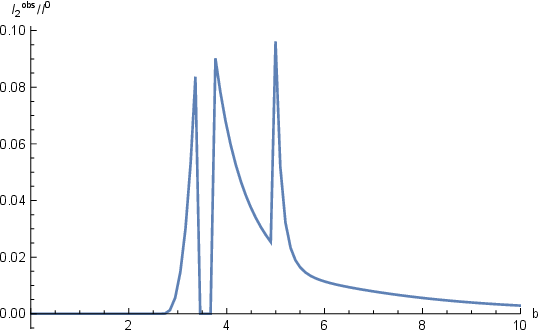}
	\end{minipage}	
      \hspace{0.5cm}
    \begin{minipage}{0.3\textwidth}
    \centering
    \vspace{-0.5cm}
		\includegraphics[scale=0.55,angle=0]{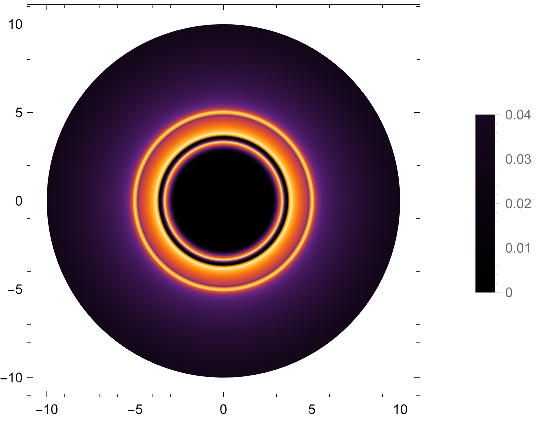}
	\end{minipage}

\vspace{2em}

\begin{minipage}{0.3\textwidth}
    \centering
		\includegraphics[scale=0.55,angle=0]{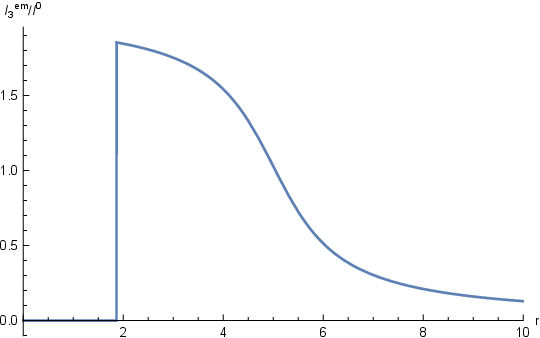}
	\end{minipage}
    \hspace{0.5cm}
    \begin{minipage}{0.3\textwidth}
    \centering
 		\includegraphics[scale=0.55,angle=0]{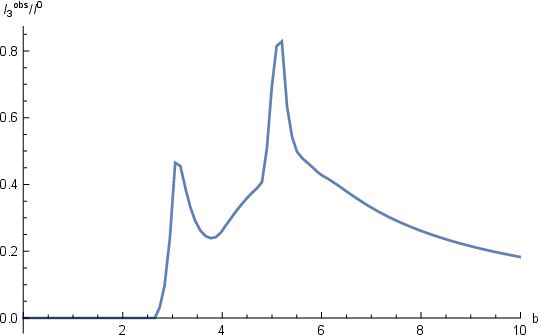}
	\end{minipage}	
      \hspace{0.5cm}
    \begin{minipage}{0.3\textwidth}
    \centering
    \vspace{-0.5cm}
		\includegraphics[scale=0.55,angle=0]{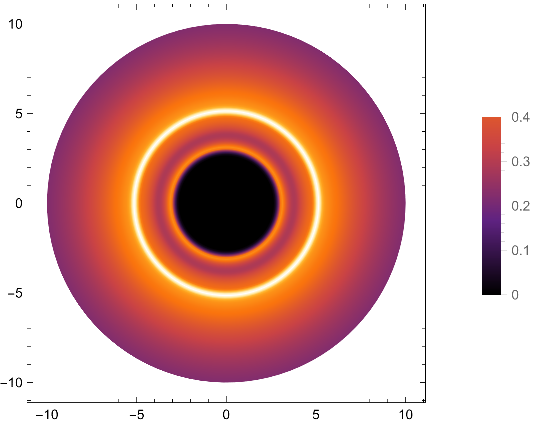}
	\end{minipage}
\caption{The observational appearance of the thin disk with different emission profiles for \( \alpha = 0, \beta = 0, q = 0.5 \) is shown from a face-on perspective. This figure presents the results of the three models corresponding to  RN BH. } 
\end{figure}

\begin{figure}[H]
\centering
	\begin{minipage}{0.3\textwidth}
    \centering
		\includegraphics[scale=0.55,angle=0]{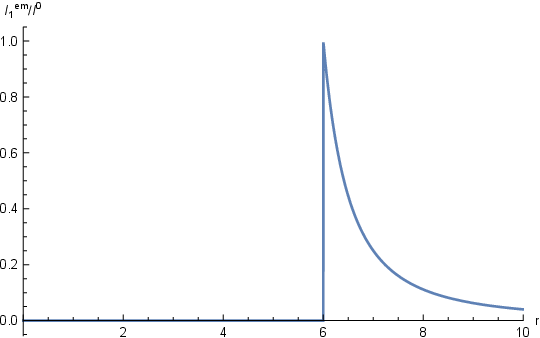}
	\end{minipage}
    \hspace{0.5cm}
    \begin{minipage}{0.3\textwidth}
   \centering
 		\includegraphics[scale=0.55,angle=0]{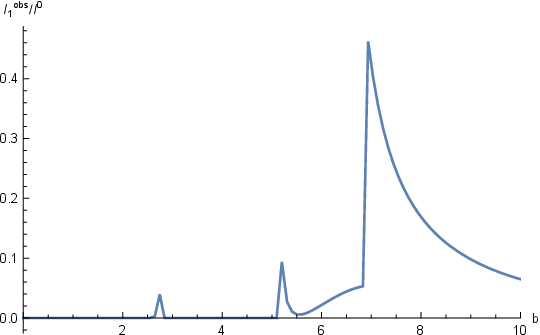}
	\end{minipage}	
    \hspace{0.5cm}
    \begin{minipage}{0.3\textwidth}
    \centering
     \vspace{-0.5cm}
		\includegraphics[scale=0.55,angle=0]{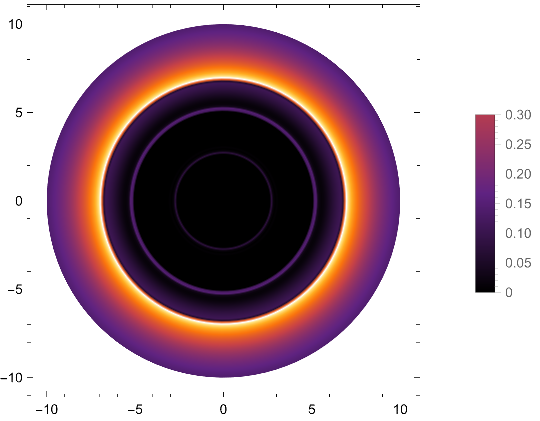}
	\end{minipage}

\vspace{2em}

\begin{minipage}{0.3\textwidth}
    \centering
		\includegraphics[scale=0.55,angle=0]{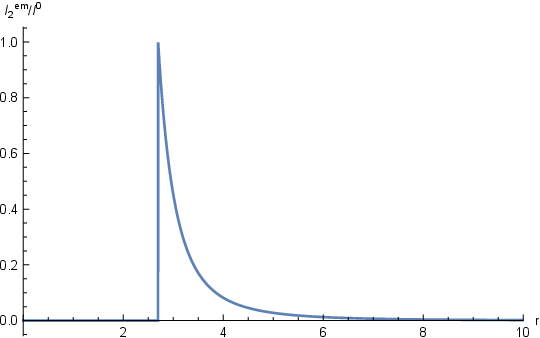}
	\end{minipage}
    \hspace{0.5cm}
    \begin{minipage}{0.3\textwidth}
    \centering
 		\includegraphics[scale=0.55,angle=0]{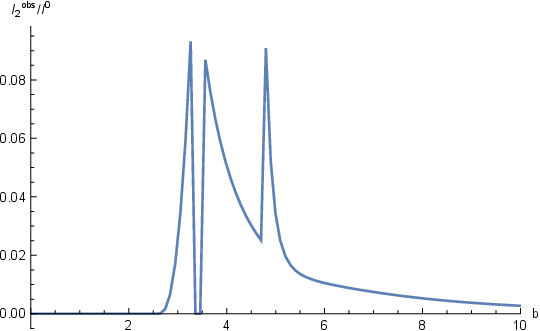}
	\end{minipage}	
     \hspace{0.5cm} 
    \begin{minipage}{0.3\textwidth}
    \centering
    \vspace{-0.5cm}
		\includegraphics[scale=0.55,angle=0]{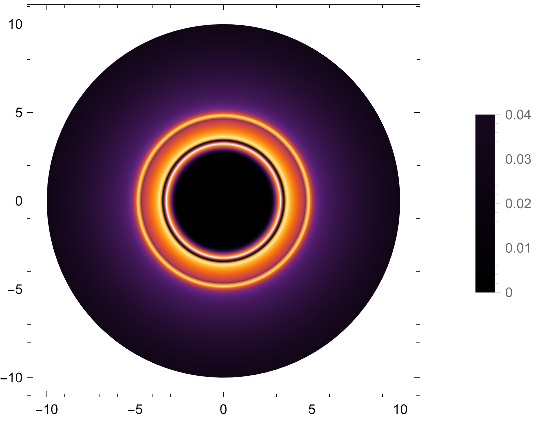}
	\end{minipage}

\vspace{2em}

\begin{minipage}{0.3\textwidth}
    \centering
		\includegraphics[scale=0.55,angle=0]{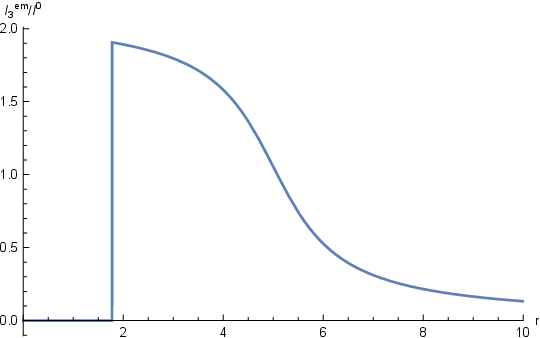}
	\end{minipage}
    \hspace{0.5cm}
    \begin{minipage}{0.3\textwidth}
    \centering
 		\includegraphics[scale=0.55,angle=0]{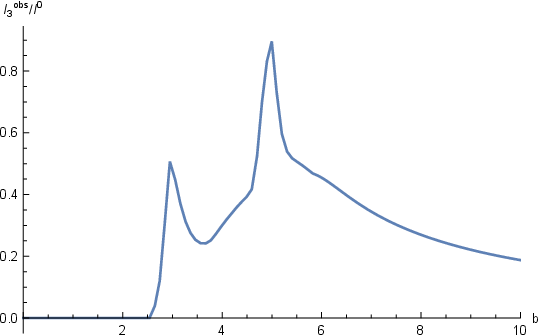}
	\end{minipage}	
      \hspace{0.5cm}
 \begin{minipage}{0.3\textwidth}
   \centering
  \vspace{-0.5cm}
		\includegraphics[scale=0.55,angle=0]{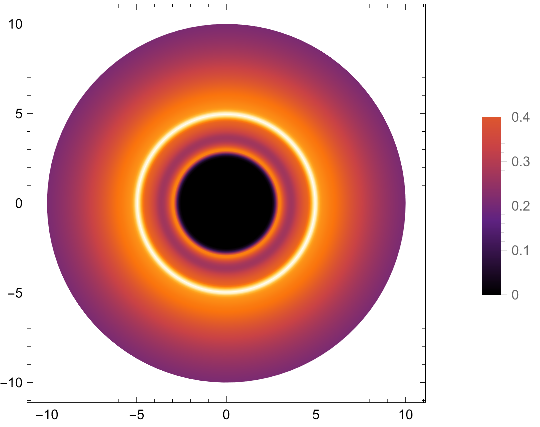}
	\end{minipage}
\caption{The observational appearance of the thin disk with different emission profiles for \( \alpha = 0.5, \beta = 0.8, q = 0.5 \) is shown from a face-on perspective. This figure presents the results of the three models corresponding to  EMS BH. } 
\end{figure}

Figures 7, 8, and 9 display the observed emission from thin accretion disks, intensity distributions, and impact parameter mappings for three BH models under different coupling parameters. In each case, the top row corresponds to the first model. The intensity profile (top-left panel) peaks near the critical impact parameter $b_c$ and gradually decreases outward. The direct image (top-middle panel) closely follows the intrinsic emission but is affected by gravitational lensing, leading to a sharp boundary and image distortion. Notably, the photon and lensing rings contribute marginally to the overall intensity due to their limited angular extent and lower peak brightness. The shadow image (top-right panel) reveals the lensing ring as a narrow, bright ring just inside the direct image, while the photon ring appears extremely faint and confined to a narrow region, visible only upon magnification. These results highlight the dominant role of direct emission in observed disk images.

% In the second model (middle row), the emission intensity reaches its maximum near the photon sphere ($r \sim r_{\mathrm{ph}}$) and gradually decreases with increasing radial distance (see the first panel in the middle row). The corresponding observed intensity profile (middle-second) also exhibits a peak primarily dominated by direct emission from the vicinity of the photon sphere, followed by a monotonic decline at larger radii. In this region, although the contributions from the photon ring and lensing ring are relatively minor, their superposition with the direct emission produces a enhancement in the total observed intensity. However, due to the strong gravitational redshift and significant depolarization experienced by photons in these rings, their observable brightness remains extremely faint and confined to a narrow angular region. As a result, despite their presence, the direct emission continues to dominate the observed intensity, a characteristic clearly visible in the two-dimensional black hole shadow image (middle-third panel).
In the second model (middle row), the emission intensity peaks near the photon sphere ($r \sim r_{\mathrm{ph}}$) and gradually decreases outward (middle-left panel). The observed intensity profile (middle-center) similarly shows a peak dominated by direct emission from this region, followed by a steady decline. Although the photon and lensing rings contribute modestly, their overlap with direct emission slightly enhances the total signal. Nonetheless, due to strong gravitational redshift and depolarization, their brightness remains faint and confined to a narrow angle. As shown in the shadow image (middle-right panel), direct emission remains the dominant observable feature.

% In the third model, illustrated in the bottom row, the intensity reaches its maximum at the event horizon ($r_+$) and gradually diminishes as the radius increases (refer to the first panel in the bottom row). Under such conditions, gravitational redshift significantly reduces the amount of radiation detectable by a distant observer. A notable distinction from the first two models lies in the fact that, here, the emissions from the photon ring and lensing ring are layered atop the direct radiation. Nonetheless, as previously noted, these ring structures contribute minimally to the overall observed brightness due to their extreme demagnetization and limited spatial extent. As a result, the direct emission remains the principal source of observed intensity, a feature that becomes evident in the two-dimensional shadow representation (see the third panel in the bottom row).
In the third model (bottom row), the emission intensity peaks near the event horizon ($r_+$) and gradually declines with increasing radius (bottom-left panel). Due to strong gravitational redshift, much of this radiation is suppressed before reaching a distant observer. Unlike the previous models, the photon and lensing ring emissions overlay the direct radiation; however, their contributions remain minimal because of severe demagnification and narrow angular extent. Consequently, the observed intensity is still primarily determined by direct emission, as clearly seen in the shadow image (bottom-right panel).

Although Fig.7, 8 and 9 show only a few highly idealized cases
of thin disk emission near a BH (viewed face on), it illustrates two key points that we believe will hold quite generally for optically thin disk
emission: (1) The emission is dominated by the direct emission, with the lensing ring emission providing only a small contribution to the total flux and the photon ring providing a negligible contribution in all cases. (2) For all three models, as we increase the coupling constant $\alpha,\beta$ and $q$, the observed intensity decreases, and it is much less than what we get for the Schwarzschild BH.

\subsection{Weak gravitational lensing and magnification of lensed image}
Let us now examine the brightness of the image in the weak field limit by using the light's deflection angle around EMS BH.  We will  find the  deflection angle by using the Gauss-Bonnet theorem (GBT), originally stated as \cite{667,668}
\begin{equation}
    \iint_D K d S+\sum_{a=1}^N \int_{\partial D_a} \kappa_{\mathrm{g}} d \ell+\sum_{a=1}^N \theta_a=2 \pi \chi(D).
\end{equation}

Consider a freely orientable $2D$ curved surface $S$ with an infinitesimal area element $dS$, which defines a domain $D$ with a Gaussian curvature $K$. The boundary of $D$ is given by $N$ components, $\partial D_{\mathrm{a}}$ (where $\mathrm{a}=1,2, \ldots, N$), along which the geodesic curvature $\kappa_{\mathrm{g}}$ is integrated over the path element $d \ell$ following a positive convention. The jump angles are denoted by $\theta_{\mathrm{a}}$. The Euler characteristic of the domain, $\chi(D)$, is equal to 1 in this case since $D$ is a non-singular region.

It was shown by Ref.\cite{666} that in spherically symmetric spacetime admitting asymptotic flatness,  the deflection angle can be written as
\begin{equation}
\hat{\alpha} = -\iint_D K \, dS.
\end{equation}
Substituting the EMS solution into the above expression and performing the weak-field approximation yields
\begin{equation}
\hat \alpha \approx \frac{4 M}{b}-\frac{3 \pi Q^2\left(1-\alpha^2+\beta\right)}{4 b^2}.
\end{equation}
    
In the context of gravitational lensing, let us consider the following lens equation, expressed in terms of the light angles $\hat{\alpha}$, $\theta$, and $\eta$ \cite{Morozova13,Bozza2008lens,Babar21a,Al_Badawi_2024}
\begin{eqnarray}\label{lenseq}
\theta D_\mathrm{s}=\eta D_\mathrm{s}+\hat{\alpha}D_\mathrm{ds}\,,
\end{eqnarray}
where $D_\mathrm{s}$ is the distance between the source and observer,  $D_\mathrm{ds}$ is the distance between the source and lens, and the angular positions of the image and the source are denoted by $\theta$ and $\eta$, respectively. From Eq.~(\ref{lenseq}), the source's angular position can be expressed as
\begin{eqnarray}\label{newlenseq}
\eta=\theta -\frac{D_\mathrm{ds}}{D_\mathrm{s}}\frac{\hat{\alpha}(\theta)}{D_\mathrm{d}}\,,
\end{eqnarray}
where $D_\mathrm{d}$ is the distance between the lens and observer. We note the use of the reduced deflection angle $\hat{\alpha}(\theta) = \xi(\theta) / b$, where $b \equiv D_d \theta$ is the impact parameter~\cite{Bozza2008lens}. The angular part, known as the Einstein radius $\theta_E$, is derived from the spacetime geometry between the source images and corresponds to the radius of the Einstein ring, provided the image forms a ring-like shape. It is given by~\cite{Morozova13}
\begin{eqnarray}
\theta_E=\sqrt{2R_s\frac{D_{ds}}{D_dD_s}}\,,
\end{eqnarray}
with $R_s=D_d \theta_E$ as the radius of the Einstein ring.

Next, we investigate the magnification of brightness, defined as the ratio of the total brightness of the lensed image ($I_{\mathrm{tot}}$) to the unlensed source brightness ($I_*$). The total magnification is given by~\cite{Babar21a,Atamurotov2022}
\begin{eqnarray}\label{magni}
\mu_\mathrm{tot}=\frac{I_\mathrm{tot}}{I_*}=\sum_k\bigg|\frac{\theta_k}{\beta}\frac{d\theta_k}{d\beta}\bigg|, \quad k=1,2, \dots , j\, .
\end{eqnarray}
For the specific case of two images, the total magnification simplifies to
\begin{eqnarray}\label{magtot}
\mu_\mathrm{tot}=\frac{x^2+2}{x\sqrt{x^2+4}}\,,
\end{eqnarray}
where $x=\eta/\theta_E$ is a dimensionless parameter.

\begin{figure}[H]
	\begin{minipage}{0.5\textwidth}
    \centering
		\includegraphics[scale=0.9,angle=0]{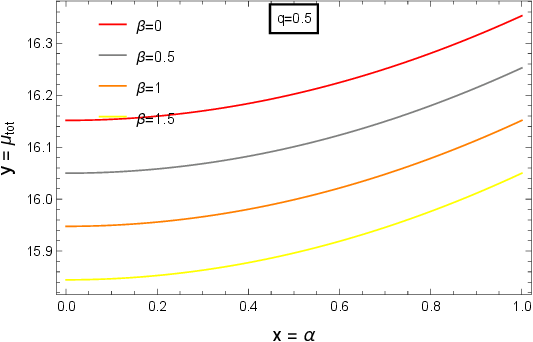}
	\end{minipage}
    \begin{minipage}{0.5\textwidth}
    \centering
		\includegraphics[scale=0.9,angle=0]{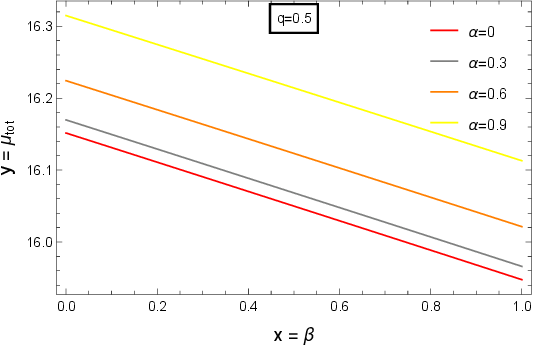}
	\end{minipage}	
    
    \begin{minipage}{0.5\textwidth}
		\includegraphics[scale=0.9,angle=0]{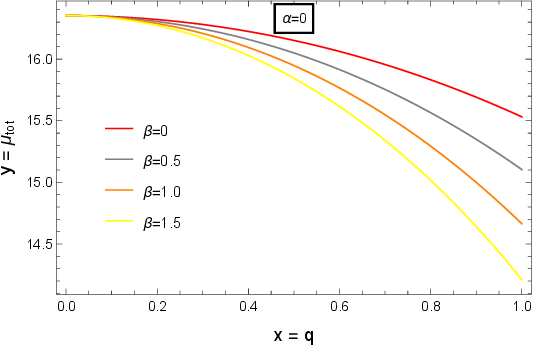}
	\end{minipage}
    \begin{minipage}{0.5\textwidth}
      \centering
		\includegraphics[scale=0.9,angle=0]{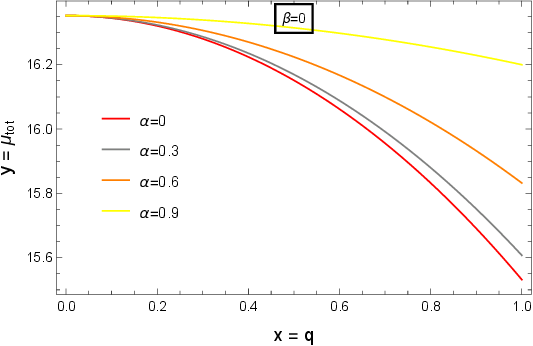}
	\end{minipage} 
\caption{Total magnification $\mu_{tot}$ plotted as a function of  the different parameters $\alpha$, $\beta$ and $q$. }\label{fig1}
\end{figure}

We conducted a numerical analysis of the total magnification for different values of the parameters $\alpha$, $\beta$, and $q$ for  EMS BH (Fig ~\ref{fig1}). Our findings show that for a fixed value of $q$, the total magnification increases with increasing $\alpha$ but decreases with increasing $\beta$. Furthermore, the total magnification decreases as the BH charge increases. We also observed a downward shift in the magnification curves for increasing values of $\beta$ when $\alpha=0$. Conversely, for $\beta=0$, we obtain the opposite trend compared to the $\alpha=0$ case.

\section{Conclusions}
In this paper, we investigated the observational characteristics of thin accretion disk surrounding the EMS BH. We first analyzed the conditions under which the BH possesses inner and outer horizons and derived the theoretical parameter range. Furthermore, we examined the geodesic equation of photons and determined the location of the photon sphere for the EMS BH. Our results indicate that as the parameter \(\alpha\) increases, the radius of the photon sphere enlarges, whereas an increase in \(\beta\) leads to a decrease in the photon sphere radius. Subsequently, employing the analytical expression for the shadow of spherically symmetric BH, we constrained the model parameters \(\alpha\) and \(\beta\) using the angular diameter distance from the EHT data within the \(1\sigma\) and \(2\sigma\) confidence intervals. Our findings reveal that the parameter \(\beta\) is well constrained, while the constraints on \(\alpha\) remain less stringent.

On the other hand, we analyzed the photon trajectories of the Schwarzschild BH, RN BH, and EMS BH. The differences among these three types of BHs are only reflected in the critical impact parameter $b_c$ (the BH shadow radius). We found that the Schwarzschild BH has the largest shadow radius, while the EMS BH has the smallest. All three types of BHs undergo a transition from direct emission to the photon ring, then to the lensing ring, and finally back to direct emission as the impact parameter $b$ increases. A similar effect has been observed using the Okyayvg Mathematical Notebook package \cite{182} in our investigation of BH shadows and rings with three toy models of thin accretion disk. Our analysis reveals that the observed intensity is primarily dominated by direct emission, while the lensing ring contributes only marginally to the total flux, and the photon ring remains negligible in all cases. Among the three types of BHs, the Schwarzschild BH exhibits the highest emission intensity, followed by the RN BH, with the EMS BH having the lowest intensity. In other words, an increase in the model parameters leads to a reduction in the disk's radiation intensity. Fianlly, when we consider weak gravitational lensing in EMS BH, our results indicate that for fixed $q$, the total magnification grows with $\alpha$ but decreases with $\beta$ and with increasing BH charge. Moreover, $\beta$ shifts the curves downward when $\alpha=0$, while for $\beta=0$ the trend reverses.
%Finally, we investigated the imaging of the accretion disk induced by the BH and found that the parameters have a negligible impact on both the direct and secondary images.

\begin{acknowledgments}
This work was supported by the National Key Research and Development Program of China (No. 2022YFA1403700), NSFC (Grants No. 12334002), Guangdong Provincial Quantum Science Strategic Initiative Grand No. SZZX2401001, the Science, Technology and Innovation Commission of Shenzhen Municipality (No. ZDSYS20190902092905285), and Center for Computational Science and Engineering at Southern University of Science and Technology.
\end{acknowledgments}

\appendix

\section{Derivations for $r_{ph}$}
For a general static, spherically symmetric spacetime, the photon sphere radius $r_{\mathrm{ph}}$ can be determined by the extremum condition of the effective potential for null geodesics. Using the metric
\begin{equation}
    ds^2 = -f(r) dt^2 + f^{-1}(r) dr^2 + C(r) \left( d\theta^2 + \sin^2\theta \, d\phi^2 \right),
\end{equation}
the condition for the photon sphere reads
\begin{equation}
    \frac{d}{dr} \left( \frac{C(r)}{f(r)} \right) = 0,
\end{equation}
or equivalently:
\begin{equation}
    f'(r) \, C(r) - f(r) \, C'(r) = 0.
\end{equation}

In our EMS black hole model, the metric functions are
\begin{equation}
\begin{cases}
f(r) = \left( 1 - \frac{b_1}{r} \right) \left( 1 - \frac{b_2}{r} \right)^{\frac{1-\alpha^2}{1+\alpha^2}} + \frac{\beta Q^2}{C(r)},\\
C(r) = r^2 \left( 1 - \frac{b_2}{r} \right)^{\frac{2\alpha^2}{1+\alpha^2}},    
\end{cases}
\end{equation}
with
\begin{equation}
\begin{cases}
b_1 = \left( 1 + \sqrt{1 - q^2(1 - \alpha^2)} \right) M, \quad \\ 
b_2 = \frac{1+\alpha^2}{1-\alpha^2} \left[ 1 - \sqrt{1 - q^2(1 - \alpha^2)} \right] M.
\end{cases}
\end{equation}

Now, for simplicity, we define
\begin{equation}
    p=\frac{1-\alpha^2}{1+\alpha^2},   \quad s=\frac{2 \alpha^2}{1+\alpha^2},
\end{equation}
then we have
\begin{equation}
\begin{cases}
C^{\prime}(r) & =2 r\left(1-\frac{b_2}{r}\right)^s+s b_2\left(1-\frac{b_2}{r}\right)^{s-1}, \\
f^{\prime}(r) & =\frac{b_1}{r^2}\left(1-\frac{b_2}{r}\right)^p+\left(1-\frac{b_1}{r}\right) p\left(1-\frac{b_2}{r}\right)^{p-1} \frac{b_2}{r^2}-\beta Q^2 \frac{C^{\prime}(r)}{C(r)^2}.
\end{cases}
\end{equation}

Substituting $f(r),f'(r),C(r)$ and $C'(r)$ into the photon sphere condition
\begin{equation}
    f'(r) \, C(r) - f(r) \, C'(r) = 0,
\end{equation}
we obtain the explicit nonlinear equation for $r_{\mathrm{ph}}$ in terms of the model parameters $\alpha, \beta, Q, M$,
\begin{equation}
\begin{aligned}
0= & {\left[\frac{b_1}{r^2}\left(1-\frac{b_2}{r}\right)^p+\left(1-\frac{b_1}{r}\right) p\left(1-\frac{b_2}{r}\right)^{p-1} \frac{b_2}{r^2}-\beta Q^2 \frac{2 r\left(1-\frac{b_2}{r}\right)^s+s b_2\left(1-\frac{b_2}{r}\right)^{s-1}}{r^4\left(1-\frac{b_2}{r}\right)^{2 s}}\right]  r^2\left(1-\frac{b_2}{r}\right)^s } \\
& -\left[\left(1-\frac{b_1}{r}\right)\left(1-\frac{b_2}{r}\right)^p+\frac{\beta Q^2}{r^2\left(1-\frac{b_2}{r}\right)^s}\right] \times\left[2 r\left(1-\frac{b_2}{r}\right)^s+s b_2\left(1-\frac{b_2}{r}\right)^{s-1}\right].
\end{aligned}
\end{equation}
 This equation is highly nonlinear and cannot be solved analytically in closed form for arbitrary parameter values. Therefore, the photon sphere radius can only be expressed implicitly.

However, once specific values of $\alpha, \beta, Q, M$ are chosen, the implicit equation can be solved numerically to yield $r_{\mathrm{ph}}$ with high precision using standard computational tools. This approach allows for direct investigation of the dependence of the photon sphere radius on the physical and coupling parameters of the EMS black hole.

\section{ Derivations for $\hat \alpha$ }
We follow the Asahi Ishihara's method \cite{666}  to calculate the deflection angle. At first, we have  
\begin{equation}
d t^2=\gamma_{i j} d x^i d x^j,
\end{equation}
where $i$ and $j$ denote 1,2 and 3 , and $\gamma_{i j}$ is often called the optical metric. Then we can have the Gaussian Curvature
\begin{equation}
K = \frac{R_{r\phi r\phi}}{g}, \quad g = \det(\gamma_{ij}) = \frac{C(r)}{f(r)^3},
\end{equation}
and the Gaussian Curvature then can be written as
\begin{equation}
K=-\frac{1}{2} \sqrt{\frac{f^3(r)}{C(r)}} \frac{\partial}{\partial r}\left(\sqrt{\frac{f^3(r)}{C(r)}} \frac{\partial}{\partial r}\left(\frac{C(r)}{f(r)}\right)\right),
\end{equation}

thus (in the weak limit $M\ll r$) 
\begin{equation}
\begin{aligned}
K \approx& \frac{1}{r^6\left(r^2-M r q^2 \alpha^2\right)^2}\left(M(3 M-2 r) r^6+M r^2 r q^2\left(r^4+3 M^3 r q^2-2 M^2 r\left(r+r q^2\right)\right) \alpha^2\right. \\
& +Q^2(2 M-r) r\left(3 r^4-4 M r^2 r q^2 \alpha^2+3 M^2 r q^4 \alpha^4\right)\left(-1+\alpha^2-\beta\right) \\
& \left.+Q^4\left(2 r^4-3 M r^2 r q^2 \alpha^2+2 M^2 r q^4 \alpha^4\right)\left(1-\alpha^2+\beta\right)^2\right).
\end{aligned}
\end{equation}

Next, we will use these equations to find the weak deflection angle using the Gauss-Bonnet theorem (GBT), originally stated as \cite{667,668}
\begin{equation}
    \iint_D K d S+\sum_{a=1}^N \int_{\partial D_a} \kappa_{\mathrm{g}} d \ell+\sum_{a=1}^N \theta_a=2 \pi \chi(D).
\end{equation}

Here, the Gaussian curvature $K$ describing the domain $D$ is a freely orientable $2 D$ curved surface $S$ with an infinitesimal area element is $d S$. The boundary of $D$ are given by $\partial D_{\text {a }}$ $(\mathrm{a}=1,2, \ldots, N)$, and the geodesic curvature $\kappa_{\mathrm{g}}$ is integrated over the path $d \ell$ along a positive convention. Also, $\theta_{\mathrm{a}}$ is the jump angle, which $\chi(D)$ is the Euler characteristic, which in our case is equal to 1 since $D$ is in a non-singular region.

It was shown by \cite{666} that in a SSS spacetime admitting asymptotic flatness,  the deflection angle can be written as
\begin{equation}
\hat{\alpha} = -\iint_D K \, dS.
\end{equation}

Taking the half-plane $\phi \in [0, \pi]$ with area element 
\begin{equation}
dS \approx r+M\left(3-\frac{a^2 q^2}{2}\right)+\frac{1}{r}\left[-\frac{3}{2} Q^2\left(1-a^2+\beta\right)+M^2\left(\frac{15}{2}-\frac{3 a^2 q^2}{2}-\frac{a^4 q^4}{8}\right)\right] dr \, d\phi,
\end{equation}
integrate over $r \in [r_0, \infty]$.

The final result is
\begin{equation}
\hat \alpha \approx \frac{4 M}{b}-\frac{3 \pi Q^2\left(1-\alpha^2+\beta\right)}{4 b^2}.
\end{equation}

\bibliographystyle{unsrt}
\bibliography{EMSbb}

\begin{thebibliography}{10}

\bibitem{124}
B.P. Abbott and Abbott.
\newblock Observation of gravitational waves from a binary black hole merger.
\newblock {\em Physical Review Letters}, 116(6), February 2016.

\bibitem{125}
Kazunori Akiyama and Alberdi.
\newblock First m87 event horizon telescope results. i. the shadow of the supermassive black hole.
\newblock {\em The Astrophysical Journal Letters}, 875(1):L1, April 2019.

\bibitem{126}
Kazunori Akiyama and Alberdi.
\newblock First m87 event horizon telescope results. ii. array and instrumentation.
\newblock {\em The Astrophysical Journal Letters}, 875(1):L2, April 2019.

\bibitem{127}
Kazunori Akiyama and Alberdi.
\newblock First m87 event horizon telescope results. iii. data processing and calibration.
\newblock {\em The Astrophysical Journal Letters}, 875(1):L3, April 2019.

\bibitem{128}
Kazunori Akiyama and Alberdi.
\newblock First m87 event horizon telescope results. iv. imaging the central supermassive black hole.
\newblock {\em The Astrophysical Journal Letters}, 875(1):L4, April 2019.

\bibitem{129}
Kazunori Akiyama and Alberdi.
\newblock First m87 event horizon telescope results. v. physical origin of the asymmetric ring.
\newblock {\em The Astrophysical Journal Letters}, 875(1):L5, April 2019.

\bibitem{130}
Kazunori Akiyama and Alberdi.
\newblock First m87 event horizon telescope results. vi. the shadow and mass of the central black hole.
\newblock {\em The Astrophysical Journal Letters}, 875(1):L6, April 2019.

\bibitem{131}
Kazunori Akiyama and Algaba.
\newblock First m87 event horizon telescope results. vii. polarization of the ring.
\newblock {\em The Astrophysical Journal Letters}, 910(1):L12, March 2021.

\bibitem{132}
Kazunori Akiyama and Algaba.
\newblock First m87 event horizon telescope results. viii. magnetic field structure near the event horizon.
\newblock {\em The Astrophysical Journal Letters}, 910(1):L13, March 2021.

\bibitem{133}
Prashant Kocherlakota and Rezzolla.
\newblock Constraints on black-hole charges with the 2017 eht observations of m87*.
\newblock {\em Physical Review D}, 103(10), May 2021.

\bibitem{134}
{Event Horizon Telescope Collaboration} and {Akiyama}.
\newblock {First Sagittarius A* Event Horizon Telescope Results. I. The Shadow of the Supermassive Black Hole in the Center of the Milky Way}.
\newblock {\em \apjl}, 930(2):L12, May 2022.

\bibitem{135}
Pedro~V.P. Cunha, Carlos~A.R. Herdeiro, and Maria~J. Rodriguez.
\newblock Does the black hole shadow probe the event horizon geometry?
\newblock {\em Physical Review D}, 97(8), April 2018.

\bibitem{136}
Rohta Takahashi.
\newblock Shapes and positions of black hole shadows in accretion disks and spin parameters of black holes.
\newblock {\em The Astrophysical Journal}, 611(2):996–1004, August 2004.

\bibitem{145}
Jose~Luis Blázquez-Salcedo, Carlos~A.R. Herdeiro, Jutta Kunz, Alexandre~M. Pombo, and Eugen Radu.
\newblock Einstein-maxwell-scalar black holes: The hot, the cold and the bald.
\newblock {\em Physics Letters B}, 806:135493, July 2020.

\bibitem{146}
Mert Okyay and Ali Övgün.
\newblock Nonlinear electrodynamics effects on the black hole shadow, deflection angle, quasinormal modes and greybody factors.
\newblock {\em JCAP}, 01:009, 2022.

\bibitem{147}
Alireza Allahyari, Mohsen Khodadi, Sunny Vagnozzi, and David~F. Mota.
\newblock Magnetically charged black holes from nonlinear electrodynamics and the event horizon telescope.
\newblock {\em JCAP}, 02:003, 2020.

\bibitem{148}
Yifan Chen, Rittick Roy, Sunny Vagnozzi, and Luca Visinelli.
\newblock Superradiant evolution of the shadow and photon ring of sgr a*.
\newblock 2022.

\bibitem{149}
Rittick Roy, Sunny Vagnozzi, and Luca Visinelli.
\newblock Superradiance evolution of black hole shadows revisited.
\newblock {\em Phys. Rev. D}, 105:083002, 2022.

\bibitem{150}
Mohsen Khodadi, Alireza Allahyari, Sunny Vagnozzi, and David~F. Mota.
\newblock Black holes with scalar hair in light of the event horizon telescope.
\newblock {\em JCAP}, 09:026, 2020.

\bibitem{151}
Sunny Vagnozzi, Rittick Roy, Yu-Dai Tsai, and Luca Visinelli.
\newblock Horizon-scale tests of gravity theories and fundamental physics from the event horizon telescope image of sagittarius a*.
\newblock 2022.

\bibitem{152}
Hui-Min Wang, Yu-Meng Xu, and Shao-Wen Wei.
\newblock Shadows of kerr-like black holes in a modified gravity theory.
\newblock {\em JCAP}, 03:046, 2019.

\bibitem{153}
Pedro V.~P. Cunha, Nelson~A. Eiró, Carlos A.~R. Herdeiro, and José P.~S. Lemos.
\newblock Lensing and shadow of a black hole surrounded by a heavy accretion disk.
\newblock {\em JCAP}, 03:035, 2020.

\bibitem{154}
Reggie~C. Pantig and Emmanuel~T. Rodulfo.
\newblock Weak deflection angle of a dirty black hole.
\newblock {\em Chin. J. Phys.}, 66:691--702, 2020.

\bibitem{155}
Reggie~C. Pantig and Emmanuel~T. Rodulfo.
\newblock Rotating dirty black hole and its shadow.
\newblock {\em Chin. J. Phys.}, 68:236--257, 2020.

\bibitem{185}
Khadije Jafarzade, Zeynab Bazyar, Sara Saghafi, and Kourosh Nozari.
\newblock Optical signatures of einstein-euler-heisenberg ads/ds black holes in the light of event horizon telescope, 2025.

\bibitem{137}
Sebastian~H Völkel, Enrico Barausse, Nicola Franchini, and Avery~E Broderick.
\newblock Eht tests of the strong-field regime of general relativity.
\newblock {\em Classical and Quantum Gravity}, 38(21):21LT01, October 2021.

\bibitem{Liu:2024brf}
Ailin Liu, Tong-Yu He, Ming Liu, Zhan-Wen Han, and Rong-Jia Yang.
\newblock {Possible signatures of higher dimension in thin accretion disk around brane world black hole}.
\newblock {\em JCAP}, 07:062, 2024.

\bibitem{Yin:2025coq}
Jia-Jun Yin, Tong-Yu He, Ming Liu, Hui-Min Fan, Bohai Chen, Zhan-Wen Han, and Rong-Jia Yang.
\newblock {Observational properties of black hole in quantum fluctuation modified gravity}.
\newblock {\em Nucl. Phys. B}, 1018:117004, 2025.

\bibitem{He:2022lrc}
Tong-Yu He, Ziqiang Cai, and Rong-Jia Yang.
\newblock {Thin accretion disks around a black hole in Einstein-Aether-scalar theory}.
\newblock {\em Eur. Phys. J. C}, 82(11):1067, 2022.

\bibitem{Wu:2024sng}
Yingdong Wu, Haiyuan Feng, and Wei-Qiang Chen.
\newblock {Thin accretion disk around black hole in Einstein{\textendash}Maxwell-scalar theory}.
\newblock {\em Eur. Phys. J. C}, 84(10):1075, 2024.

\bibitem{Feng:2024iqj}
Haiyuan Feng, Rong-Jia Yang, and Wei-Qiang Chen.
\newblock {Thin accretion disk and shadow of Kerr{\textendash}Sen black hole in Einstein{\textendash}Maxwell-dilaton{\textendash}axion gravity}.
\newblock {\em Astropart. Phys.}, 166:103075, 2025.

\bibitem{Feng:2023iha}
Haiyuan Feng, Yingdong Wu, Rong-Jia Yang, and Leonardo Modesto.
\newblock {Choked accretion onto Kerr-Sen black holes in Einstein-Maxwell-dilaton-axion gravity}.
\newblock {\em Phys. Rev. D}, 109(6):063014, 2024.

\bibitem{Feng:2022bst}
Haiyuan Feng, Miao Li, Gui-Rong Liang, and Rong-Jia Yang.
\newblock {Adiabatic accretion onto black holes in Einstein-Maxwell-scalar theory}.
\newblock {\em JCAP}, 04(04):027, 2022.

\bibitem{138}
Clifford~M. Will.
\newblock The confrontation between general relativity and experiment.
\newblock {\em Living Reviews in Relativity}, 17(1), June 2014.

\bibitem{139}
Emanuele Berti, Alessandra Buonanno, and Clifford~M. Will.
\newblock Estimating spinning binary parameters and testing alternative theories of gravity with lisa.
\newblock {\em Physical Review D}, 71(8), April 2005.

\bibitem{140}
DAMIEN~A. EASSON.
\newblock Modified gravitational theories and cosmic acceleration.
\newblock {\em International Journal of Modern Physics A}, 19(31):5343–5350, December 2004.

\bibitem{141}
SHIN’ICHI NOJIRI and SERGEI~D. ODINTSOV.
\newblock Introduction to modified gravity and gravitational alternative for dark energy.
\newblock {\em International Journal of Geometric Methods in Modern Physics}, 04(01):115–145, February 2007.

\bibitem{142}
MARK TRODDEN.
\newblock Cosmic acceleration and modified gravity.
\newblock {\em International Journal of Modern Physics D}, 16(12a):2065–2074, December 2007.

\bibitem{180}
Shao-Wen Wei and Yu-Xiao Liu.
\newblock Observing the shadow of einstein-maxwell-dilaton-axion black hole.
\newblock {\em Journal of Cosmology and Astroparticle Physics}, 2013(11):063–063, November 2013.

\bibitem{Gibbons:1988}
G.~W. Gibbons and K.~Maeda.
\newblock Black holes and membranes in higher-dimensional theories with dilaton fields.
\newblock {\em Nucl. Phys. B}, 298:741, 1988.

\bibitem{Garfinkle:1990qj}
David Garfinkle, Gary~T. Horowitz, and Andrew Strominger.
\newblock {Charged black holes in string theory}.
\newblock {\em Phys. Rev. D}, 43:3140, 1991.
\newblock [Erratum: Phys.Rev.D 45, 3888 (1992)].

\bibitem{Polchinski:1998}
J.~Polchinski.
\newblock {\em String Theory}, volume I and II.
\newblock Cambridge University Press, 1998.

\bibitem{Becker:2006}
K.~Becker, M.~Becker, and J.~H. Schwarz.
\newblock {\em String Theory and M-Theory: A Modern Introduction}.
\newblock Cambridge University Press, 2006.

\bibitem{Cremmer:1979}
E.~Cremmer and B.~Julia.
\newblock The so(8) supergravity.
\newblock {\em Nucl. Phys. B}, 159:141, 1979.

\bibitem{deWit:1982}
B.~de~Wit and H.~Nicolai.
\newblock N=8 supergravity.
\newblock {\em Nucl. Phys. B}, 208:323, 1982.

\bibitem{Duff:1996}
M.~J. Duff, J.~T. Liu, and J.~Rahmfeld.
\newblock Four-dimensional string-string-string triality.
\newblock {\em Nucl. Phys. B}, 459:125, 1996.

\bibitem{belkhadria_2025}
Zakaria Belkhadria and Salvatore Mignemi.
\newblock A new model of spontaneous scalarization of black holes induced by curvature and matter, 2025.

\bibitem{Qiu_2021}
Jianhui Qiu.
\newblock Slowly rotating black holes in the novel einstein–maxwell-scalar theory.
\newblock {\em The European Physical Journal C}, 81(12), December 2021.

\bibitem{143}
Carlos~A.R. Herdeiro, Eugen Radu, Nicolas Sanchis-Gual, and José~A. Font.
\newblock Spontaneous scalarization of charged black holes.
\newblock {\em Physical Review Letters}, 121(10), September 2018.

\bibitem{144}
Pedro G~S Fernandes, Carlos A~R Herdeiro, Alexandre~M Pombo, Eugen Radu, and Nicolas Sanchis-Gual.
\newblock Spontaneous scalarisation of charged black holes: coupling dependence and dynamical features.
\newblock {\em Classical and Quantum Gravity}, 36(13):134002, June 2019.

\bibitem{189}
Zakaria Belkhadria and Alexandre~M. Pombo.
\newblock Mixed scalarization of charged black holes: From spontaneous to nonlinear scalarization.
\newblock {\em Physical Review D}, 110(4), August 2024.

\bibitem{102}
I.~Z. Fisher.
\newblock {Scalar mesostatic field with regard for gravitational effects}.
\newblock {\em Zh. Eksp. Teor. Fiz.}, 18:636--640, 1948.

\bibitem{103}
Th~Kaluza.
\newblock Zum unit{\"a}tsproblem der physik.
\newblock {\em Sitzungsber. Preuss. Akad. Wiss. Berlin (Math. Phys.)}, 1921(arXiv: 1803.08616):966--972, 1921.

\bibitem{104}
Peter Van~Nieuwenhuizen.
\newblock Supergravity.
\newblock {\em Physics Reports}, 68(4):189--398, 1981.

\bibitem{105}
Jérôme Martin and Jun’ichi Yokoyama.
\newblock Generation of large scale magnetic fields in single-field inflation.
\newblock {\em Journal of Cosmology and Astroparticle Physics}, 2008(01):025, January 2008.

\bibitem{3}
Chang~Jun Gao and Shuang~Nan Zhang.
\newblock Higher-dimensional dilaton black holes with cosmological constant.
\newblock {\em Physics Letters B}, 605(1–2):185–189, January 2005.

\bibitem{4}
S.~Hajkhalili and A.~Sheykhi.
\newblock Topological dyonic dilaton black holes in ads spaces.
\newblock {\em Phys. Rev. D}, 99:024028, Jan 2019.

\bibitem{5}
S.~H. Hendi, A.~Sheykhi, and M.~H. Dehghani.
\newblock Thermodynamics of higher dimensional topological charged ads black branes in dilaton gravity.
\newblock {\em The European Physical Journal C}, 70(3):703–712, November 2010.

\bibitem{6}
S.~Hajkhalili and A.~Sheykhi.
\newblock {Asymptotically (A)dS dilaton black holes with nonlinear electrodynamics}.
\newblock {\em Int. J. Mod. Phys. D}, 27(07):1850075, 2018.

\bibitem{7}
Ahmad Sheykhi.
\newblock {Charged rotating dilaton black strings in AdS spaces}.
\newblock {\em Phys. Rev. D}, 78:064055, 2008.

\bibitem{8}
Ryo Yamazaki and Daisuke Ida.
\newblock {Black holes in three-dimensional Einstein-Born-Infeld dilaton theory}.
\newblock {\em Phys. Rev. D}, 64:024009, 2001.

\bibitem{1}
Gary~W Gibbons and Kei-ichi Maeda.
\newblock Black holes and membranes in higher-dimensional theories with dilaton fields.
\newblock {\em Nuclear Physics B}, 298(4):741--775, 1988.

\bibitem{2}
David Garfinkle, Gary~T. Horowitz, and Andrew Strominger.
\newblock {Charged black holes in string theory}.
\newblock {\em Phys. Rev. D}, 43:3140, 1991.
\newblock [Erratum: Phys.Rev.D 45, 3888 (1992)].

\bibitem{Yu_2021}
Shuang Yu, Jianhui Qiu, and Changjun Gao.
\newblock Constructing black holes in einstein–maxwell-scalar theory.
\newblock {\em Classical and Quantum Gravity}, 38(10):105006, April 2021.

\bibitem{9}
Chang~Jun Gao and Shuang~Nan Zhang.
\newblock Dilaton black holes in the de sitter or anti–de sitter universe.
\newblock {\em Physical Review D}, 70(12), December 2004.

\bibitem{177}
Akhil Uniyal, Reggie~C. Pantig, and Ali Övgün.
\newblock Probing a non-linear electrodynamics black hole with thin accretion disk, shadow, and deflection angle with m87* and sgr a* from eht.
\newblock {\em Physics of the Dark Universe}, 40:101178, May 2023.

\bibitem{176}
Volker Perlick and Oleg~Yu. Tsupko.
\newblock Calculating black hole shadows: Review of analytical studies.
\newblock {\em Physics Reports}, 947:1–39, February 2022.

\bibitem{184}
Volker Perlick, Oleg~Yu. Tsupko, and Gennady~S. Bisnovatyi-Kogan.
\newblock Influence of a plasma on the shadow of a spherically symmetric black hole.
\newblock {\em Physical Review D}, 92(10), November 2015.

\bibitem{111}
Yosuke Mizuno, Ziri Younsi, Christian~M. Fromm, Oliver Porth, Mariafelicia De~Laurentis, Hector Olivares, Heino Falcke, Michael Kramer, and Luciano Rezzolla.
\newblock The current ability to test theories of gravity with black hole shadows.
\newblock {\em Nature Astronomy}, 2(7):585–590, April 2018.

\bibitem{112}
Kocherlakota and Shan-Shan Zhao.
\newblock Constraints on black-hole charges with the 2017 eht observations of m87*.
\newblock {\em Physical Review D}, 103(10), May 2021.

\bibitem{110}
Sunny Vagnozzi, Rittick Roy, Yu-Dai Tsai, Luca Visinelli, Misba Afrin, Alireza Allahyari, Parth Bambhaniya, Dipanjan Dey, Sushant~G Ghosh, Pankaj~S Joshi, Kimet Jusufi, Mohsen Khodadi, Rahul~Kumar Walia, Ali Övgün, and Cosimo Bambi.
\newblock Horizon-scale tests of gravity theories and fundamental physics from the event horizon telescope image of sagittarius a*.
\newblock {\em Classical and Quantum Gravity}, 40(16):165007, July 2023.

\bibitem{113}
Kazunori~Milagros Akiyama, Shuo Zhang, and Lucy Ziurys.
\newblock First m87 event horizon telescope results. i. the shadow of the supermassive black hole.
\newblock {\em The Astrophysical Journal Letters}, 875(1):L1, April 2019.

\bibitem{114}
Event Horizon~Telescope Collaboration, Akiyama, and Milagros Zeballos.
\newblock First sagittarius a* event horizon telescope results. i. the shadow of the supermassive black hole in the center of the milky way.
\newblock {\em The Astrophysical Journal Letters}, 930(2):L12, May 2022.

\bibitem{115}
Deng Wang.
\newblock Shaving the hair of black hole with sagittarius a* from event horizon telescope, 2022.

\bibitem{116}
Cosimo Bambi, Katherine Freese, Sunny Vagnozzi, and Luca Visinelli.
\newblock Testing the rotational nature of the supermassive object m87* from the circularity and size of its first image.
\newblock {\em Physical Review D}, 100(4), August 2019.

\bibitem{178}
Jun Peng, Minyong Guo, and Xing-Hui Feng.
\newblock Influence of quantum correction on black hole shadows, photon rings, and lensing rings*.
\newblock {\em Chinese Physics C}, 45(8):085103, August 2021.

\bibitem{179}
Jinsong Yang, Cong Zhang, and Yongge Ma.
\newblock Shadow and stability of quantum-corrected black holes.
\newblock {\em The European Physical Journal C}, 83(7), July 2023.

\bibitem{120}
Samuel~E. Gralla, Daniel~E. Holz, and Robert~M. Wald.
\newblock Black hole shadows, photon rings, and lensing rings.
\newblock {\em Physical Review D}, 100(2), July 2019.

\bibitem{667}
W.~Klingenberg and D.~Hoffman.
\newblock {\em A Course in Differential Geometry}.
\newblock Graduate Texts in Mathematics. Springer New York, 2013.

\bibitem{668}
M.P. do~Carmo.
\newblock {\em Differential Geometry of Curves and Surfaces: Revised and Updated Second Edition}.
\newblock Dover Books on Mathematics. Dover Publications, 2016.

\bibitem{666}
Asahi Ishihara, Yusuke Suzuki, Toshiaki Ono, Takao Kitamura, and Hideki Asada.
\newblock Gravitational bending angle of light for finite distance and the gauss-bonnet theorem.
\newblock {\em Physical Review D}, 94(8), October 2016.

\bibitem{Morozova13}
V.~S. Morozova, B.~J. Ahmedov, and A.~A. Tursunov.
\newblock {Gravitational lensing by a rotating massive object in a plasma}.
\newblock {\em Astrophys. Space Sci.}, 346(2):513--520, 2013.

\bibitem{Bozza2008lens}
V.~Bozza.
\newblock Comparison of approximate gravitational lens equations and a proposal for an improved new one.
\newblock {\em Phys. Rev. D}, 78:103005, Nov 2008.

\bibitem{Babar21a}
Gulmina~Zaman Babar, Farruh Atamurotov, and Abdullah~Zaman Babar.
\newblock Gravitational lensing in 4-d einstein-gauss-bonnet gravity in the presence of plasma, 2021.

\bibitem{Al_Badawi_2024}
Ahmad Al-Badawi, Mirzabek Alloqulov, Sanjar Shaymatov, and Bobomurat Ahmedov.
\newblock Shadows and weak gravitational lensing for black holes within einstein-maxwell-scalar theory*.
\newblock {\em Chinese Physics C}, 48(9):095105, September 2024.

\bibitem{Atamurotov2022}
Farruh Atamurotov, Ahmadjon Abdujabbarov, and Javlon Rayimbaev.
\newblock Weak gravitational lensing schwarzschild-mog black hole in plasma.
\newblock {\em The European Physical Journal C}, 81:118, 2021.

\bibitem{182}
Mert Okyay and Ali Ovgun.
\newblock {Nonlinear electrodynamics effects on the black hole shadow, deflection angle, quasinormal modes and greybody factors}.
\newblock {\em JCAP}, 01(01):009, 2022.

\end{thebibliography}
\end{document}